\documentclass[11pt]{article}
\pdfoutput=1
\usepackage{amsmath,amsfonts,amssymb,color}
\usepackage{cite,url}
\usepackage{graphicx}
\usepackage[utf8]{inputenc} 
\usepackage[bottom]{footmisc}
\usepackage{hyperref}

\allowdisplaybreaks

\def\msk2lam{$m_s\approx 5.5$~TeV, $m_a\approx 4.2$~TeV}

\def\smallSM{{\rm{\scriptscriptstyle SM}}}
\def\smallBSM{{\rm{\scriptscriptstyle BSM}}}

\def\beq{\begin{equation}}
\def\eeq{\end{equation}}
\def\bea{\begin{eqnarray}}
\def\eea{\end{eqnarray}}
\def\nn{\nonumber}

\def\MM{M^2}
\def\Theavy{\widetilde T}
\def\aem{\alpha_{\rm em}}
\def\mHq{m^2_H}
\def\mAq{m^2_A}
\def\mHpq{m^2_{H^\pm}}

\def\sdbe{s_{2\beta}}
\def\cdbe{c_{2\beta}}
\def\sbe{s_{\beta}}
\def\cbe{c_{\beta}}

\def\sq2{\sqrt{2}}

\def\msbar{\overline{\rm MS}}

\def\smallIPI{{\scriptscriptstyle {\rm 1PI}}}

\long\def\symbolfootnote[#1]#2{\begingroup%
\def\thefootnote{\fnsymbol{footnote}}\footnote[#1]{#2}\endgroup}

\makeatletter
\newcommand{\vast}{\bBigg@{3}}
\makeatother

\begin{document}

\begin{titlepage}

\begin{center}

\vspace{1cm}

{\LARGE \bf On the two-loop BSM corrections to $h\longrightarrow
  \gamma\gamma$ }\\[3mm]

{\LARGE \bf in the aligned THDM} 

\vspace{1cm}

{\Large Giuseppe Degrassi$^{\,a}$ and Pietro~Slavich$^{\,b}$}

\vspace*{1cm}

{\sl ${}^a$ 
Dipartimento di Matematica e Fisica, Università di Roma Tre, 

and INFN, Sezione di Roma Tre, I-00146 Rome, Italy.}
\vspace*{2mm}\\{\sl ${}^b$
   Sorbonne Université, CNRS,
  Laboratoire de Physique Th\'eorique et Hautes Energies, 
 
  LPTHE, F-75005, Paris, France.}
\end{center}
\symbolfootnote[0]{{\tt e-mail:}}
\symbolfootnote[0]{{\tt giuseppe.degrassi@uniroma3.it}}
\symbolfootnote[0]{{\tt slavich@lpthe.jussieu.fr}}

\vspace{0.7cm}

\abstract{We compute the two-loop BSM contributions to the
  $h\longrightarrow \gamma\gamma$ decay width in the aligned THDM. We
  adopt the simplifying assumptions of vanishing EW gauge couplings
  and vanishing mass of the SM-like Higgs boson, which allow us to
  exploit a low-energy theorem connecting the $h\gamma\gamma$
  amplitude to the derivative of the photon self-energy w.r.t.~the
  Higgs field. We briefly discuss the numerical impact of the
  newly-computed contributions, showing that they may be required for
  a precise determination of $\Gamma[h\rightarrow \gamma \gamma]$ in
  scenarios where the quartic Higgs couplings are large.}

\vfill

\end{titlepage}


\setcounter{footnote}{0}

\section{Introduction}
\label{sec:intro}

The discovery of a Higgs boson with mass around $125$~GeV and
properties compatible with the predictions of the Standard Model
(SM)~\cite{CMS:2012qbp, ATLAS:2012yve, ATLAS:2015yey, ATLAS:2016neq},
combined with the negative (so far) results of the searches for
additional new particles at the LHC, point to scenarios with at least
a mild hierarchy between the electroweak (EW) scale and the scale of
beyond-the-SM (BSM) physics. This said, the existence of new particles
with masses around or even below the TeV scale, which could still be
discovered in the current or future runs of the LHC, is not
conclusively ruled out. This is especially the case if those new
particles are colorless, and there is some mechanism that forbids or
at least suppresses their mixing with the SM-like Higgs boson.

The Two-Higgs-Doublet Model (THDM) is one of the simplest and
best-studied extensions of the SM (for reviews see, e.g.,
refs.~\cite{Gunion:1989we,Aoki:2009ha,Branco:2011iw}). In the
CP-conserving versions of the model, the Higgs sector includes five
physical states: two CP-even scalars, $h$ and $H$; one CP-odd scalar,
$A$; and two charged scalars, $H^\pm$. As discussed, e.g., in
ref.~\cite{Gunion:2002zf}, the so-called ``alignment'' condition -- in
which one of the CP-even scalars has SM-like couplings to fermions and
gauge bosons -- can be realized through decoupling, when all of the
other Higgs bosons are much heavier, or without decoupling, when a
specific configuration of parameters in the Lagrangian suppresses the
mixing between the SM-like scalar and the other CP-even scalar. If the
THDM is embedded in a more-complicated extension of the SM that
predicts the values of the quartic Higgs couplings, as is the case in
supersymmetric models, the alignment condition can arise at the tree
level from an underlying symmetry (see, e.g.,
ref.~\cite{Antoniadis:2006uj}), or it can result from cancellations
between the tree-level couplings and their radiative corrections (see,
e.g., ref.~\cite{Carena:2013ooa}).  In contrast, when the THDM is
treated as a stand-alone extension of the SM, the alignment condition
can be enforced ``from the bottom up'', based on the empirical
observation that the couplings of the $125$-GeV Higgs boson appear to
be essentially SM-like.

Beyond the requirement that they allow for a scalar with mass around
$125$~GeV and SM-like couplings to fermions and gauge bosons, the
quartic Higgs couplings of the THDM are subject to a number of
experimental constraints from EW precision observables and flavor
physics, as well as theory-driven constraints from perturbative
unitarity and the stability of the scalar potential. Nevertheless,
couplings of ${\cal O}(1$--$10)$ are still allowed by all constraints
(see, e.g., refs.~\cite{Arco:2020ucn, Arco:2022xum}), and may even be
favored (see, e.g., ref.~\cite{Bahl:2022xzi}) if the THDM is to
accommodate recent experimental anomalies such as the new CDF
measurement of the $W$ mass~\cite{CDF:2022hxs}. Couplings in this
range may induce sizable radiative corrections to the THDM predictions
for physical observables, up to the point where one might wonder
whether, in any given calculation, the uncomputed higher-order effects
spoil the accuracy of the prediction. This has motivated a number of
recent studies in which radiative corrections involving the quartic
Higgs couplings of the THDM have been computed at the two-loop
level. In particular, the two-loop corrections to the $\rho$ parameter
have been computed in refs.~\cite{Hessenberger:2016atw,
  Hessenberger:2022tcx}, various effects of the two-loop corrections
to the scalar mass matrices have been examined in
ref.~\cite{Braathen:2017izn}, and the two-loop corrections to the
trilinear self-coupling of the SM-like Higgs boson, $\lambda_{hhh}$,
have been computed in refs.~\cite{Braathen:2019pxr,
  Braathen:2019zoh}. In all cases it was found that the two-loop
corrections can significantly modify the one-loop predictions, and
should be taken into account for a precise determination of the
considered observable.

In this paper we compute the dominant two-loop corrections to the
decay width for the process $h\longrightarrow \gamma \gamma$ in the
aligned THDM.  Since the signal strength for this channel is currently
measured with an accuracy of about
$6\%$~\cite{ParticleDataGroup:2022pth}, the requirement that the BSM
contributions do not spoil the agreement with the theoretical
prediction can put significant constraints on the parameter space of
the THDM (see, e.g., ref.~\cite{Arco:2022jrt}). Once again, the
possible presence of couplings of ${\cal O}(1$--$10)$ in the THDM
Lagrangian motivates the calculation of $\Gamma[h\rightarrow \gamma
  \gamma]$ beyond the leading order (LO), which, for this observable,
means beyond the one-loop level.

In the calculation of the two-loop BSM contributions to
$\Gamma[h\rightarrow \gamma \gamma]$ we adopt the same simplifying
assumptions as in the calculation of the $\rho$ parameter in
refs.~\cite{Hessenberger:2016atw, Hessenberger:2022tcx}. In
particular, we restrict our calculation to the CP-conserving THDM in
the alignment limit; we work in the so-called ``gaugeless limit'' of
vanishing EW gauge couplings, considering only the corrections that
depend on the quartic Higgs couplings and possibly on the top Yukawa
coupling; finally, we treat the mass of the SM-like scalar $h$ as
negligible w.r.t.~the masses of the BSM scalars, $H$, $A$ and
$H^{\pm}$, and of the top quark. In order to obtain compact formulas
for the two-loop corrections to the decay width, we make use of a
low-energy theorem (LET) which connects them to the derivative of the
photon self-energy w.r.t.~the vacuum expectation value (vev) of the
SM-like Higgs field~\cite{Shifman:1979eb, Kniehl:1995tn}. However, we
also cross-check our result via a direct calculation of the $h \gamma
\gamma$ amplitude. We note that care must be devoted to the definition
of the alignment and vanishing-Higgs-mass conditions beyond LO, as
well as to the avoidance of infrared (IR)-divergent contributions from
diagrams involving massless particles.

The rest of the article is organized as follows: in
section~\ref{sec:THDM} we fix our notation for the Higgs sector of the
THDM and discuss the renormalization of the scalar masses and mixing;
in section~\ref{sec:twoloop} we outline our calculation of the
dominant two-loop corrections to the decay width for $h\longrightarrow
\gamma \gamma$; in section~\ref{sec:numerics} we briefly discuss the
numerical impact of the newly-computed corrections;
section~\ref{sec:conclusions} contains our conclusions; finally, two
appendices collect explicit formulas for the one-loop self-energies
and tadpoles of the Higgs bosons and for the BSM part of the two-loop
self-energy of the photon.

\section{The Higgs sector of the aligned THDM}
\label{sec:THDM}

We start this section by describing the tree-level scalar potential
and the Higgs mass spectrum of the THDM in the alignment limit. Note
that we do not need to distinguish between different THDM ``types''
according to the form of their Higgs--fermion interactions, because in
our calculation of the two-loop corrections to $\Gamma[h\rightarrow
  \gamma \gamma]$ we neglect all Yukawa couplings except the one of
the top quark. We follow up by discussing the renormalization of the
Higgs sector of the aligned THDM, in a ``naive'' approach that is
justified by the simplifying assumptions adopted in our calculation.

\subsection{The scalar potential, masses and mixing at the tree level}
\label{sec:tree}
We consider a version of the THDM where flavor-changing
neutral-current interactions are forbidden at the tree level by a $Z_2$
symmetry, softly broken by an off-diagonal mass term. In the so-called
standard basis where this $Z_2$ symmetry applies, the scalar potential
can be parametrized as
\bea
V_0 &=& m_{11}^2 \Phi_1^\dagger \Phi_1 ~+~ m_{22}^2 \Phi_2^\dagger \Phi_2
~-~ m_{12}^2\,\left(\Phi_1^\dagger \Phi_2 ~+~ {\rm h.c.}\right)
\,+~ \frac{\lambda_1}{2}\,\left(\Phi_1^\dagger \Phi_1\right)^2
\,+~ \frac{\lambda_2}{2}\,\left(\Phi_2^\dagger \Phi_2\right)^2\nn\\[2mm]
&& +~ \lambda_3\,\Phi_1^\dagger \Phi_1\,\Phi_2^\dagger \Phi_2
~+~ \lambda_4 \,\Phi_1^\dagger \Phi_2\,\Phi_2^\dagger \Phi_1
~+~ \frac{\lambda_5}{2}\,\left[\left(\Phi_1^\dagger \Phi_2\right)^2 +~ {\rm h.c.}\right]~,
\label{eq:Vstd}
\eea
where all of the masses and quartic couplings are assumed to be real
to ensure CP conservation.  We decompose the two $SU(2)$ doublets as
\beq
\Phi_k ~=~ \frac{1}{\sqrt2} \left(\!\begin{array}{c} \sqrt2\,\phi^+_k \\ 
v_k + \phi^0_k + i\,a_k  \end{array}\!\right)~~~~~(k=1,2)~,
\eeq
where the two (real) vevs are related by $v_1^2+v_2^2=v^2$, with
$v\approx 246$~GeV, and we define $\tan\beta \equiv v_2/v_1$. The
minimum conditions for the scalar potential can be used to replace the
mass parameters $m_{11}^2$ and $m_{22}^2$ with combinations of the
remaining parameters in eq.~(\ref{eq:Vstd}) and the vevs:
\bea
\label{eq:min1}
m_{11}^2 &=& m_{12}^2\,\tan\beta - \frac{v^2}{2}\,\left(\lambda_1\,\cbe^2
+ \lambda_{345}\,\sbe^2\right)~,\\[1mm]
\label{eq:min2}
m_{22}^2 &=& m_{12}^2\,\cot\beta - \frac{v^2}{2}\,\left(\lambda_2\,\sbe^2
+ \lambda_{345}\,\cbe^2\right)~,
\eea
where we introduced the shortcuts $c_\theta\equiv \cos\theta$ and
$s_\theta\equiv \sin\theta$ for a generic angle $\theta$, and defined
$\lambda_{345}\equiv \lambda_3+\lambda_4+\lambda_5$. The mass matrices
for the pseudoscalar and charged components of the two doublets are
diagonalized by the angle $\beta$:
\beq
\left(\!\begin{array}{c} G^0 \\ A \end{array}\!\right) ~=~
R(\beta)\, \left(\!\begin{array}{c}
a_1\\ a_2 \end{array}\!\right)~,~~~~~
\left(\!\begin{array}{c} G^+ \\ H^+ \end{array}\!\right) ~=~
R(\beta)\,\left(\!\begin{array}{c}
\phi_1^+\\ \phi_2^+ \end{array}\!\right)~,
\eeq
where we defined
\beq
R(\theta) ~\equiv~\left(\!\begin{array}{rr} c_\theta&\!s_\theta\\-s_\theta
&\! c_\theta \end{array}\!\right)~,
\eeq
and using the minimum conditions from eqs.~(\ref{eq:min1}) and
(\ref{eq:min2}) we get the tree-level masses
\beq
m^2_{G^0} = m^2_{G^\pm}=0~,~~~
m_A^2 = \MM \,-\, \lambda_5\,v^2~,~~~
m_{H^\pm}^2 = \MM \,-\, \frac12(\lambda_4+\lambda_5)\,v^2~,
\eeq
where we defined $\MM\equiv m_{12}^2/(\sbe\cbe)$. The mass matrix for
the neutral scalar components of the two doublets is instead
diagonalized by an angle $\alpha$:
\beq
\label{eq:rotstd}
\left(\!\begin{array}{c} H \\ h \end{array}\!\right) ~=~
R(\alpha)\, \left(\!\begin{array}{c}
\phi^0_1\\ \phi^0_2 \end{array}\!\right)~,
\eeq
and the alignment condition in which the lighter mass eigenstate $h$
has SM-like couplings to fermions and gauge bosons corresponds to
$\alpha=\beta - \pi/2$.
To discuss this condition and its eventual renormalization, it is
convenient to rotate the original Higgs doublets to the so-called
Higgs basis:
\beq
\label{eq:hb}
\left(\!\begin{array}{c} \Phi_{\smallSM} \\ \Phi_{\smallBSM} \end{array}\!\right)
~=~ R(\beta)\,\left(\!\begin{array}{c}
\Phi_1 \\ \Phi_2 \end{array}\!\right)~,
\eeq
in which one of the doublets develops the full SM-like vev $v$ and the
other has vanishing vev:
\beq
\Phi_\smallSM ~=~ \left(\!\begin{array}{c} G^+ \\ \frac{1}{\sqrt2}
(v + \phi^0_\smallSM + i\,G^0)  \end{array}\!\right)~,~~~~~
\Phi_\smallBSM ~=~ \left(\!\begin{array}{c} H^+ \\ \frac{1}{\sqrt2}
(\phi^0_\smallBSM + i\,A)  \end{array}\!\right)~.
\eeq
The scalar potential in the Higgs basis becomes
\bea
V_0 &=& M_{11}^2\, \Phi_\smallSM^\dagger \Phi_\smallSM
~+~ M_{22}^2 \,\Phi_\smallBSM^\dagger \Phi_\smallBSM
~-~ M_{12}^2\,\left(\Phi_\smallSM^\dagger \Phi_\smallBSM ~+~ {\rm h.c.}\right)
\nn\\[2mm]
&&
+~ \frac{\Lambda_1}{2}\,\left(\Phi_\smallSM^\dagger \Phi_\smallSM\right)^2
\,+~ \frac{\Lambda_2}{2}\,\left(\Phi_\smallBSM^\dagger \Phi_\smallBSM\right)^2
\nn\\[2mm]
&&
+~ \Lambda_3\,\left(\Phi_\smallSM^\dagger \Phi_\smallSM\right)
\left(\Phi_\smallBSM^\dagger \Phi_\smallBSM\right)
+~ \Lambda_4\,\left(\Phi_\smallSM^\dagger \Phi_\smallBSM\right)
\left(\Phi_\smallBSM^\dagger \Phi_\smallSM\right)\nn\\[2mm]
&&
+~ \left[\,
\frac{\Lambda_5}{2}\,\left(\Phi_\smallSM^\dagger \Phi_\smallBSM\right)^2
+\, \left( \Lambda_6 \,\Phi_\smallSM^\dagger \Phi_\smallSM
\,+\,\Lambda_7\,\Phi_\smallBSM^\dagger \Phi_\smallBSM \right)
\,
\Phi_\smallSM^\dagger \Phi_\smallBSM ~+~{\rm h.c.}\right]~.
\label{eq:Vhb}
\eea
The general relations between the parameters in eq.~(\ref{eq:Vhb}) and
the analogous parameters in the standard basis, eq.~(\ref{eq:Vstd}),
are given, e.g., in the appendix of ref.~\cite{Davidson:2005cw}. We
list here the ones that will be relevant to the discussion that
follows, specialized to the case of the $Z_2$-symmetric THDM:
\bea
\label{eq:M11}
M_{11}^2 &=& m_{11}^2\,\cbe^2 + m_{22}^2\,\sbe^2 -  m_{12}^2\,\sdbe~,\\[1mm]
\label{eq:M22}
M_{22}^2 &=& m_{11}^2\,\sbe^2 + m_{22}^2\,\cbe^2 +  m_{12}^2\,\sdbe~,\\[1mm]
\label{eq:M12}
M_{12}^2 &=& \frac12\,\left(m_{11}^2 - m_{22}^2\right)\,\sdbe
+  m_{12}^2\,\cdbe~,\\[1mm]
\label{eq:L1}
\Lambda_1 &=& \lambda_1 \,\cbe^4 \,+\, \lambda_2\,\sbe^4
+ \frac12\,\lambda_{345}\,\sdbe^2~,\\[1mm]
\label{eq:L6}
\Lambda_6 &=& -\frac12\,\sdbe \,
\left(\lambda_1\,\cbe^2-\lambda_2\,\sbe^2-\lambda_{345}\,\cdbe\right)~.
\eea
The minimum conditions for the scalar potential become
\beq
M_{11}^2 ~=~ -\frac{\Lambda_1}{2}\,v^2~,~~~~
M_{12}^2 ~=~ \frac{\Lambda_6}{2}\,v^2~,
\label{eq:minhb}
\eeq
and the mass parameter for the BSM doublet in eq.~(\ref{eq:M22}) can
be rewritten as
\beq
M_{22}^2 ~=~ \MM
\,-\,\frac12\,\left(\Lambda_1
\,+\,2\,\cot2\beta\,\Lambda_6\right) \,v^2~.
\label{eq:M22hbtree}
\eeq
In the Higgs basis the tree-level mass matrices for the pseudoscalar
and charged components of the two doublets are already diagonal, and
the tree-level mass matrix for the neutral scalar components is given
by
\beq
V~\supset~ \frac12\,
\left( \phi^0_\smallSM~~~\phi^0_\smallBSM\right)
     ~{\cal M}_0^2~ \left(\!\begin{array}{c}
\phi^0_\smallSM\\ \phi^0_\smallBSM \end{array}\!\right)~,~~~~~~~~~
{\cal M}_0^2 =
\left(\!\begin{array}{cc} \Lambda_1\,v^2& \Lambda_6\,v^2\\
\Lambda_6\,v^2
&\MM
\,+\,\widetilde\Lambda\,v^2
\end{array}\right)~,
\label{eq:Mhbtree}
\eeq
where
$\widetilde\Lambda\,=\,(\lambda_1+\lambda_2-2\lambda_{345})\,\sdbe^2/4$\,.
It is then clear that, at the tree level, the alignment condition
corresponds to $\Lambda_6=0$. When that is the case, the masses of the
lighter and heavier neutral scalar reduce to
\beq
m_h^2~\rightarrow~ \Lambda_1\,v^2~,~~~~~~
m_H^2~\rightarrow~ \MM\,+\,\widetilde\Lambda\,v^2~,
\eeq
where we used arrows to indicate that the relations holds only in the
alignment limit. Similarly, eq.~(\ref{eq:M22hbtree}) reduces to
\beq
M_{22}^2~\rightarrow~ \MM\,-\, \frac{m_h^2}2~,
\eeq
and two of the quartic couplings of the standard basis can be traded
for combinations of the remaining parameters:
\beq
\label{eq:l1l2tree}
\lambda_1~\rightarrow~
- \lambda_{345}\,\tan^2\beta\,+\,\frac{m_h^2}{v^2\,\cbe^2}~,~~~~~
\lambda_2~\rightarrow~
- \lambda_{345}\,\cot^2\beta\,+\,\frac{m_h^2}{v^2\,\sbe^2}~,
\eeq
which implies $\mHq\rightarrow \MM + m_h^2 - \lambda_{345}\,v^2$.  We
remark that the approximation of vanishing $m_h$, which we will adopt
in section~\ref{sec:twoloop} to simplify our two-loop results, has to
be understood here as $\Lambda_1 \approx 0$ rather than $v\approx 0$,
i.e., it amounts to a condition on the couplings entering
eq.~(\ref{eq:L1}).  Finally, we note that the combination of
eqs.~(\ref{eq:rotstd}) and (\ref{eq:hb}) implies
\beq
\label{eq:rotAmB}
\left(\!\begin{array}{c} H \\ h \end{array}\!\right) ~=~
R(\alpha-\beta)\, \left(\!\begin{array}{c}
\phi^0_\smallSM\\ \phi^0_\smallBSM \end{array}\!\right)~,
\eeq
thus, when $\alpha=\beta-\pi/2$ we get $h\rightarrow\phi^0_\smallSM$
and $H \rightarrow -\phi^0_\smallBSM$.

\subsection{Mass and mixing renormalization}
\label{sec:renorm}

The calculation of two-loop corrections to $\Gamma[h\rightarrow \gamma
  \gamma]$ requires one-loop definitions for the parameters entering
the LO prediction, which is itself at the one-loop level.  The full
one-loop renormalization of the Higgs sector of the THDM has been
extensively studied in the literature~\cite{Kanemura:2004mg,
  Krause:2016oke, Altenkamp:2017ldc, Kanemura:2017wtm}, and it
involves a number of subtleties concerning the possible gauge
dependence of the renormalized mixing angles. However, the simplifying
assumptions that we adopt in our calculation (namely, alignment
condition, vanishing SM-like Higgs mass $m_h$, and vanishing EW gauge
couplings) allow us to bypass most of the complications discussed in
those earlier studies. What we will ultimately need in the computation
of the two-loop BSM corrections to $\Gamma[h\rightarrow \gamma
  \gamma]$ is the renormalization of the charged-Higgs mass
$m_{H^\pm}$ and of the parameters $v$, $m_{12}^2$ and $\beta$ (the
latter two make up $\MM$), taking care that the alignment and
vanishing-$m_h$ conditions hold at the perturbative level considered
in our calculation.

\bigskip

Beyond the tree level, the minimum conditions of the scalar potential
in eq.~(\ref{eq:minhb}) become
\beq
M_{11}^2 ~=~ -\frac{\Lambda_1}{2}\,v^2
~-~ \frac{T_{\phi_\smallSM^0}}{v}~,~~~~~~~
M_{12}^2 ~=~ \frac{\Lambda_6}{2}\,v^2
~+~ \frac{T_{\phi_\smallBSM^0}}{v}~,
\label{eq:minhbloop}
\eeq
where $M^2_{11}$, $M^2_{12}$, $\Lambda_1$, $\Lambda_6$ and $v$ are now
interpreted as $\msbar$-renormalized parameters at some scale $Q$.
Note that eqs.~(\ref{eq:M11})--(\ref{eq:L6}) imply that the masses and
quartic couplings in the standard basis, as well as the angle $\beta$,
are also interpreted as $\msbar$-renormalized parameters. The
quantities $T_{\varphi}$ in eq.~(\ref{eq:minhbloop}) denote the finite
parts of the one-loop tadpole diagrams\,\footnote{Decomposing the
  effective potential as $V_0 + \Delta V$, we also have $T_\varphi
  \,=\, d \Delta V/d\varphi$.}  for the fields $\varphi =
(\phi_\smallSM^0,\phi_\smallBSM^0$). The relation between $M_{22}^2$
and $\MM$ in eq.~(\ref{eq:M22hbtree}) becomes in turn
\beq
M_{22}^2 ~=~ \MM
\,-\,\frac12\,\left(\Lambda_1
\,+\,2\,\cot2\beta\,\Lambda_6\right) \,v^2
~-~\frac{\Theavy}{v}~,
\label{eq:M22hbloop}
\eeq
where we define
\beq
\Theavy ~=~ T_{\phi_\smallSM^0} + 2\,\cot 2\beta\,T_{\phi_\smallBSM^0}
~\rightarrow~  T_{h} - 2\,\cot 2\beta\,T_{H}~.
\eeq
Again, we use the arrow to indicate that the second equality holds
only in the alignment limit (note the sign flip when going from
$\phi_\smallBSM^0$ to $H$).

The mass matrix for the neutral scalar components of the doublets in
the Higgs basis also receives radiative corrections, i.e., ${\cal
  M}^2(p^2) \,=\, {\cal M}_0^2 + \Delta {\cal M}^2(p^2)$, where $p^2$ is
the external momentum. The tree-level part ${\cal M}_0^2$, now
expressed in terms of $\msbar$-renormalized parameters, is given in
eq.~(\ref{eq:Mhbtree}), and
\beq
\Delta {\cal M}^2(p^2) ~=~
\left(\!\begin{array}{cc}
\Pi_{\phi_\smallSM^0\phi_\smallSM^0}(p^2) &
\Pi_{\phi_\smallSM^0\phi_\smallBSM^0}(p^2)\\[2mm]
  \Pi_{\phi_\smallSM^0\phi_\smallBSM^0}(p^2)&
 \Pi_{\phi_\smallBSM^0\phi_\smallBSM^0}(p^2) 
\end{array}\right)
~-~ \frac{1}{v}\,
\left(\!\begin{array}{cc}
T_{\phi_\smallSM^0} & T_{\phi_\smallBSM^0}\\[2mm]
T_{\phi_\smallBSM^0} & \Theavy   
\end{array}\right)~,
\label{eq:DMhb}
\eeq
where $\Pi_{\varphi\varphi^\prime}(p^2)$ are the finite parts of the
$2\!\times\!2$ one-loop self-energy matrix for the neutral scalars. We
can now implement the alignment condition beyond the tree level by
requiring that ${\cal M}^2_{12}(p^2)$ vanish for $p^2=m_h^2\,$, i.e.,
for the external momentum that is relevant to the calculation of the
$h\longrightarrow \gamma\gamma$ amplitude. This also implies that
${\cal M}^2_{11}(m_h^2)$ corresponds to the squared pole mass $M_h^2$
of the SM-like Higgs boson. We therefore require\,\footnote{Here and
thereafter, we use $M_\varphi^2$ to denote the squared pole mass of a
scalar $\varphi$, but keep using $m_\varphi^2$ when the precise
definition of the mass amounts to a higher-order effect.}
\bea
\label{eq:renL1}
M_h^2 &=&\Lambda_1\,v^2 ~+~ \Pi_{hh}(m_h^2)
\,~-~ \frac{T_h}{v}~,\\[1mm]
\label{eq:renL6}
0 &=& \Lambda_6\,v^2 ~-~ \Pi_{hH}(m_h^2)
~+~ \frac{T_H}{v}~.
\eea
%
%
These conditions can now be used to remove $\Lambda_1$ and $\Lambda_6$
from eq.~(\ref{eq:M22hbloop}), which becomes
\beq
M_{22}^2 ~=~
 \MM
~-~\frac{M_h^2}{2}
~+~ \frac12\biggr( \Pi_{hh}(m_h^2)
\,-\, 2\,\cot2\beta \, \Pi_{hH}(m_h^2)\biggr)
\,-~ \frac32\,\frac{\Theavy}{v}~.
\label{eq:M22hbloopplus}
\eeq
If we then consider the {\em pole} mass of the SM-like Higgs boson to
be negligible w.r.t.~the BSM Higgs masses,
eq.~(\ref{eq:M22hbloopplus}) reduces to
\beq
M_{22}^2 ~=~
 \MM
~+~ \frac12\biggr( \Pi_{hh}(0)
\,-\, 2\,\cot2\beta \, \Pi_{hH}(0)\biggr)
\,-~ \frac32\,\frac{\Theavy}{v}~.
\label{eq:M22hbloopMh0}
\eeq

We now need to discuss the renormalization of the charged-Higgs
mass. It is possible to define two different running masses, depending
on whether or not the minimum conditions of the scalar potential have
been used to replace $M_{22}^2$ with $\MM$:
\bea
\label{eq:mHphat}
\widehat m_{H^\pm}^2 &=&  M_{22}^2 \,+\,\frac12\,\left(\Lambda_1
\,+\,2\,\cot2\beta\,\Lambda_6\right) \,v^2 \,-\,
\frac12(\lambda_4+\lambda_5)\,v^2 ~,\\[1mm]
\label{eq:mHptil}
\widetilde m_{H^\pm}^2 &=& \MM\,-\,
\frac12(\lambda_4+\lambda_5)\,v^2~.
\eea
At the tree level these two definitions would coincide due to
eq.~(\ref{eq:M22hbtree}), but at the one-loop level they differ by the
tadpole contribution entering eq.~(\ref{eq:M22hbloop}):
\beq
\label{eq:mHphattil}
\widehat m_{H^\pm}^2~=~\widetilde m_{H^\pm}^2 \,-\, \frac{\Theavy}{v}~.
\eeq
Finally, the two definitions of the running mass of the charged Higgs
boson are related to the corresponding pole mass by
\bea
\label{eq:polehat}
M_{H^\pm}^2
&=&\widehat m_{H^\pm}^2  ~+~ {\rm Re}\,\Pi_{H^+H^-}(m_{H^\pm}^2)\\[1mm]
\label{eq:poletil}
&=&\widetilde m_{H^\pm}^2  
~+~ {\rm Re}\,\Pi_{H^+H^-}(m_{H^\pm}^2)~-~ \frac{\Theavy}{v}~.
\eea

Explicit formulas for the Higgs tadpoles and self-energies under the
approximations relevant to our two-loop calculation are collected in
the appendix A.

\bigskip

To conclude this section, it might be useful to compare our approach
to the renormalization of the scalar mixing with the approaches of
refs.~\cite{Kanemura:2004mg, Krause:2016oke, Altenkamp:2017ldc,
  Kanemura:2017wtm}, which are not restricted to the alignment
limit. In those approaches, the amplitude for a process that involves
an external SM-like scalar $h$ receives counterterm contributions from
the renormalization of the angles $\alpha$ and $\beta$ that enter the
couplings of $h$ prior to taking the limit $\alpha \rightarrow \beta -
\pi/2$, as well as from the off-diagonal wave-function renormalization
(WFR) of the Higgs scalars. In contrast, since our calculation is
restricted to the alignment limit, we choose not to introduce an angle
$\alpha$ at all, and our eq.~(\ref{eq:renL6}) is equivalent to the
requirement that the contribution of the off-diagonal WFR to the
mixing of $h$ and $H$ be cancelled by a small but non-vanishing
tree-level contribution.
Once again, we stress that the gaugeless limit is what allows us to
sidestep the complications related to the gauge-dependence of the
renormalization conditions that were discussed in
refs.~\cite{Krause:2016oke, Altenkamp:2017ldc, Kanemura:2017wtm}.

\vfill
\newpage

\section{Leading two-loop contributions to
  $\Gamma[h\rightarrow \gamma \gamma]$}
\label{sec:twoloop}

We now discuss our calculation of the dominant two-loop corrections to
$\Gamma[h\rightarrow \gamma \gamma]$ in the aligned (and
CP-conserving) THDM. As mentioned in the previous sections, we adopt
the same simplifying assumptions as in the calculation of the $\rho$
parameter in refs.~\cite{Hessenberger:2016atw, Hessenberger:2022tcx},
working in the limit of vanishing EW gauge couplings, neglecting all
Yukawa couplings except the top one, and treating the mass of the
SM-like Higgs boson as negligible w.r.t.~the masses of the BSM Higgs
bosons and of the top quark. The fact that we restrict our calculation
to the alignment limit of the THDM allows us to neatly separate the
contributions involving the BSM Higgs bosons from those that are in
common with the SM. We do not need to compute the latter as they can
already be found in the literature, see refs.~\cite{Zheng:1990qa,
  Djouadi:1990aj, Dawson:1992cy, Melnikov:1993tj, Djouadi:1993ji,
  Inoue:1994jq, Fleischer:2004vb} for the QCD corrections and
refs.~\cite{Liao:1996td, Djouadi:1997rj, Fugel:2004ug,
  Degrassi:2005mc} for the EW corrections involving the top
quark.\footnote{The remaining EW corrections have also been computed,
see refs.~\cite{Aglietti:2004nj, Aglietti:2004ki, Passarino:2007fp,
  Actis:2008ts}.}

\bigskip

The partial width for the $h\longrightarrow \gamma \gamma$ decay can
be written as
\beq
\label{eq:width}
\Gamma(h\rightarrow \gamma\gamma) ~=~
\frac{G_\mu\,\aem^2 M_h^3}{128\,\sqrt2\,\pi^3}
\,\left|{\cal P}_h^{1\ell}+{\cal P}_h^{2\ell}\right|^2~,
\eeq
where $\aem$ is the electromagnetic coupling and $G_\mu$ is the Fermi
constant, which is proportional to $v^{-2}$ at the tree level. ${\cal
  P}_h^{1\ell}$ and ${\cal P}_h^{2\ell}$ denote the one- and two-loop
$h\gamma\gamma$ amplitudes, respectively. The latter can be further
decomposed as
\beq
\label{eq:splitP2l}
{\cal P}_h^{2\ell} ~=~ {\cal P}_h^{2\ell,\,\smallIPI} \,+\,
\delta {\cal P}_h^{1\ell} \,+\,K_r\, {\cal P}_h^{1\ell}~,
\eeq
where: ${\cal P}_h^{2\ell,\,\smallIPI}$ denotes the genuine two-loop
part, in which we include the one-particle-irreducible (1PI)
contributions as well as the $\msbar$ counterterm contributions;
$\delta {\cal P}_h^{1\ell}$ stems from renormalization-scheme choices
for the parameters entering ${\cal P}_h^{1\ell}$; the additional
correction factor $K_r$ accounts for the diagonal WFR of the external
Higgs field and for the connection between $v$ and $G_\mu$ beyond the
tree level. As mentioned above, we will focus on the calculation of
the BSM part of the two-loop amplitude, which we denote as ${\cal
  P}_h^{2\ell,\,\smallBSM}$.

\bigskip

In the approximation of vanishing external momentum for the
$h\gamma\gamma$ amplitude (i.e., vanishing mass for the SM-like Higgs
boson), the LET of refs.~\cite{Shifman:1979eb, Kniehl:1995tn} allows
us to write
\beq
\label{eq:LET}
    {\cal P}_h^{1\ell} ~=~ \frac{2\pi\,v}{\aem}~\frac{d\,
      \Pi_{\gamma\gamma}^{1\ell} (0)}{dv}~,
    ~~~~~~~~~
    {\cal P}_h^{2\ell,\,\smallIPI} ~=~ \frac{2\pi\,v}{\aem}\,\frac{d\,
      \Pi_{\gamma\gamma}^{2\ell}(0)}{dv}~,
\eeq
where $\Pi_{\gamma\gamma}(0)$ denotes the transverse part of the
dimensionless self-energy of the photon at vanishing external
momentum. At the one-loop level, it reads
\bea
\Pi_{\gamma\gamma}^{1\ell}(0)&=&
\Pi_{\gamma\gamma}^{1\ell,\,H^\pm}(0)
~+~\Pi_{\gamma\gamma}^{1\ell,\,t}(0)
~+~\Pi_{\gamma\gamma}^{1\ell,\,W}(0)\nonumber\\[2mm]
&=&
\frac{\aem}{4\pi}\,\left( \frac 13 \,\ln  \frac{\,\widehat m_{H^\pm}^2}{Q^2}
~+~ \frac 43 \,Q_t^2\,N_c \,\ln \frac{m_{t}^2}{Q^2}
~-~ 7 \,\ln \frac{m_{W}^2}{Q^2} \,+\,\frac 23 ~ \right)~,
\label{eq:Pigaga1l}
\eea
where $N_c=3$ is a color factor, $Q_t=2/3$ is the electric charge of
the top quark, and we omitted all other fermionic contributions
because our calculation of the $h\gamma\gamma$ amplitude neglects the
corresponding Yukawa couplings. The contribution of the gauge sector,
$\Pi_{\gamma\gamma}^{1\ell,\,W}(0)$, is in fact gauge dependent, and
only when computed in the unitary gauge or in the background-field
gauge (or using the pinch technique) can it be directly connected to
${\cal P}_h^{1\ell}$ through eq.~(\ref{eq:LET}).  We also remark that
our choice to express the charged-Higgs contribution in terms of the
running mass $\widehat m^2_{H^\pm}$, see eq.~(\ref{eq:mHphat}), will
affect the determination of $\delta {\cal P}_h^{1\ell}$.

\bigskip

We computed the contributions to the transverse part of the photon
self-energy from two-loop diagrams that involve the BSM Higgs bosons,
which we denote as $\Pi_{\gamma\gamma}^{2\ell,\,\smallBSM}(0)$, with
the help of {\tt FeynArts}~\cite{Hahn:2000kx}. We performed our
calculation in the unitary gauge, including also the contributions
from diagrams that involve gauge bosons together with the BSM Higgs
bosons. When the self-energy is Taylor-expanded in powers of the
external momentum $p^2$, the zeroth-order term of the expansion
vanishes as a consequence of gauge invariance, while the first-order
term corresponds to $\Pi_{\gamma\gamma}^{2\ell,\,\smallBSM}(0)$. We
evaluated the two-loop vacuum integrals using the results of
ref.~\cite{Davydychev:1992mt}. Only after performing the momentum
expansion did we take the ``gaugeless limit'' of vanishing EW gauge
couplings, except for an overall factor $\aem$ from the couplings of
the external photons. We remark that this procedure avoids
complications related to the presence of massless would-be-Goldstone
bosons, which would have affected our calculation if we had tried to
impose the gaugeless limit from the start.
For what concerns the SM-like Higgs boson, we took the limit of
vanishing mass after the momentum expansion. Finally, in the diagrams
that involve the bottom quark, the top quark, and the charged Higgs
boson, we set the bottom mass directly to zero before the momentum
expansion. As a cross-check, we recomputed those diagrams by means of
an asymptotic expansion analogous to the one described in section 3 of
ref.~\cite{Degrassi:2010eu}, and found the same result. Explicit
formulas for $\Pi_{\gamma\gamma}^{2\ell,\,\smallBSM}(0)$ as function
of the BSM Higgs masses, the top mass, $\MM$ and $\beta$ can be found
in the appendix B.

\bigskip

In the alignment limit, the derivative of a given field-dependent
quantity w.r.t.~the SM-like Higgs field $h$ can be replaced by the
derivative w.r.t.~$v$, see eq.~(\ref{eq:LET}). When computing the
derivative of $\Pi_{\gamma\gamma}^{2\ell,\,\smallBSM}(0)$, it is
sufficient to consider the tree-level dependence on $v$ of the masses
of the particles circulating in the loops. Since at the tree level
$m_t^2 = y_t^2\, s_\beta^2\,v^2/2$, where $y_t$ is the top Yukawa
coupling, and $m_\Phi^2 = \MM \,+\, \tilde\lambda_\Phi\,v^2$, where
$\Phi = (H,A,H^\pm)$ and $\tilde \lambda_\Phi$ are combinations of
quartic Higgs couplings, the use of the chain rule for the derivative
w.r.t.~$v$ leads to
\beq
\label{eq:catenone}
\frac{d}{dv}~=~
\frac{\partial}{\partial v} ~+~ \frac{2}{v}\,\left[  
  m_t^2\,\frac{\partial}{\partial m_t^2}
  \,+\,(\mHq-\MM)\,\frac{\partial}{\partial \mHq}
  \,+\,(\mAq-\MM)\,\frac{\partial}{\partial \mAq}
  \,+\,(\mHpq-\MM)\,\frac{\partial}{\partial \mHpq}\,
  \right]~.
\eeq  
It is now straightforward to compute the BSM part of ${\cal
  P}_h^{2\ell,\,\smallIPI}$ by applying the operator in
eq.~(\ref{eq:catenone}) to the explicit expression for
$\Pi_{\gamma\gamma}^{2\ell,\,\smallBSM}(0)$ given in the appendix
B. Since the result is lengthy and not particularly illuminating, we
refrain from putting it in print and we make it available on request
in electronic form.

\vfill
\newpage

The second contribution to the two-loop amplitude ${\cal P}_h^{2\ell}$
in eq.~(\ref{eq:splitP2l}) arises from the renormalization of the
parameters entering the one-loop amplitude ${\cal P}_h^{1\ell}$. The
SM part of the latter is

\bea
    {\cal P}_h^{1\ell,\,\smallSM}
    &=&
    \frac{2\pi\,v}{\aem}
    \,\frac{d}{dv}\,
    \left[\Pi_{\gamma\gamma}^{1\ell,\,t}(0)
      \,+\,\Pi_{\gamma\gamma}^{1\ell,\,W}(0)\right]\nn\\[2mm]
    &=&
    \frac{2\,Q_t^2\,N_c\,v}{3\,m_t^2}\,
    \frac{d\,m_t^2}{dv}
    ~-~\frac{7\,v}{2\,m_W^2}\,
    \frac{d\,m_W^2}{dv}\nn\\[1mm]
    &=&
    \frac43 \,Q_t^2\,N_c ~-~ 7~,
    \label{eq:P1lSM}
    \eea
where we used $m_W^2 = g^2\,v^2/4$, with $g$ being the $SU(2)$ gauge
coupling. Eq.~(\ref{eq:P1lSM}) shows that, in the limit of vanishing
Higgs mass, the SM part of ${\cal P}_h^{1\ell}$ does not involve any
parameters for which we need to define a renormalization scheme. For
the BSM part, since we expressed the charged-Higgs contribution to the
one-loop self-energy of the photon in terms of $\widehat m^2_{H^\pm}$,
the dependence on $v$ is given by eq.~(\ref{eq:mHphat}). We thus
obtain
\bea
    {\cal P}_h^{1\ell,\,\smallBSM}
    &=&
    \frac{2\pi\,v}{\aem}
\,\frac{d}{dv}\, \Pi_{\gamma\gamma}^{1\ell,\,H^\pm}(0)\nn\\[2mm]
&=&
\frac{v}{6\,\widehat m_{H^\pm}^2}\,
\frac{d\,\widehat m_{H^\pm}^2}{dv}\nn\\[1mm]
&=&
\label{eq:dP1hdv}
\frac{1}{3}\,\left(1-\frac{M_{22}^2}{\widehat m_{H^\pm}^2}\right)~.
\eea
However, we opt to re-express the one-loop amplitude in terms of the
parameter $\MM$ and of the squared pole mass of the charged Higgs
boson, $M^2_{H^\pm}$. Hence
\beq
{\cal P}_h^{1\ell,\,\smallBSM} ~=~ 
\frac{1}{3}\,\left(1-\frac{\MM}{M_{H^\pm}^2}\right)
~+~ \delta {\cal P}_h^{1\ell,\,\smallBSM}~,
\label{eq:P1lBSM}
\eeq
where the shift $\delta {\cal P}_h^{1\ell,\,\smallBSM}$, which becomes
part of ${\cal P}_h^{2\ell,\,\smallBSM}$, is determined by
eqs.~(\ref{eq:M22hbloopMh0}) and (\ref{eq:polehat}):
\beq
\label{eq:dPh1l}
\delta {\cal P}_h^{1\ell,\,\smallBSM}
~=~
- \frac{1}{3\,m_{H^\pm}^2}\,
\left[\,\frac12\biggr( \Pi_{hh}(0)
\,-\, 2\,\cot2\beta \, \Pi_{hH}(0)\biggr)
-\, \frac32\,\frac{\Theavy}{v}
\, \right]
\,-~  \frac{\MM}{3\,m_{H^\pm}^4}~
{\rm Re}\,\Pi_{H^+H^-}(m_{H^\pm}^2)~.
\eeq
It might now be instructive to consider an alternative derivation of
$\delta {\cal P}_h^{1\ell,\,\smallBSM}$. By means of
eq.~(\ref{eq:mHphattil}), the charged-Higgs contribution to the
one-loop self-energy of the photon can be re-expressed as
\beq
\label{eq:shiftPigaga}
\Pi_{\gamma\gamma}^{1\ell,\,H^\pm}(0) ~=~
\frac{\aem}{12\pi}\,\left(\ln\widetilde m_{H^\pm}^2
\,-\,\frac{\Theavy}{v\,m^2_{H^\pm}}\right)~.
\eeq
Hence, a derivation analogous to the one of eq.~(\ref{eq:dP1hdv})
leads to
\beq
{\cal P}_h^{1\ell,\,\smallBSM} ~=~
\frac{1}{3}\,\left(1-\frac{\MM}
  {\widetilde m_{H^\pm}^2}\right)
  ~-~\frac{v}{6}~\frac{d}{dv}~\frac{\Theavy}{v\,m_{H^\pm}^2}~,
\eeq  
and by means of eq.~(\ref{eq:poletil}) we obtain
\beq
\label{eq:dPh1lbis}
\delta {\cal P}_h^{1\ell,\,\smallBSM}
~=~
-\, \frac{v}{6}~\frac{d}{dv}~\frac{\Theavy}{v\,m_{H^\pm}^2}
~-~\frac{\MM}{3\,m_{H^\pm}^4}~
\left({\rm Re}\,\Pi_{H^+H^-}(m_{H^\pm}^2) - \frac{\Theavy}{v}\right)~.
\eeq
The equivalence between the two expressions for $\delta {\cal
  P}_h^{1\ell,\,\smallBSM}$, eqs.~(\ref{eq:dPh1l}) and
(\ref{eq:dPh1lbis}), relies on the identity
\beq
\frac{d}{dv}\,\biggr(T_h - 2\,\cot2\beta \,T_H\biggr) ~=~
\Pi_{hh}(0)
\,-\, 2\,\cot2\beta \, \Pi_{hH}(0)~,
\eeq
which can be checked with the formulas for tadpoles and self-energies
listed in the appendix A.

\bigskip

The third contribution to the two-loop amplitude ${\cal P}_h^{2\ell}$
in eq.~(\ref{eq:splitP2l}), i.e., $K_r\,{\cal P}_h^{1\ell}$, arises
from the diagonal WFR of the external Higgs field and from the
renormalization of the parameter $v$ that is factored out of the
amplitude in eq.~(\ref{eq:LET}):
\beq
\label{eq:Kr}
K_r ~=~ \frac12 \left(\delta Z_{hh} - \frac{\delta v^2}{v^2}\right)~,
\eeq
where
\beq    
\delta Z_{hh} ~=~ 
\left.\frac{d\, \Pi_{hh}(p^2)}{dp^2}\right|_{p^2=0}~,~~~~~~~~~~~~
\frac{\delta v^2}{v^2} ~=~ \frac{\Pi_{WW}(0)}{m_W^2}~,
\eeq
$\Pi_{WW}(0)$ being the transverse part of the $W$-boson self-energy
at zero external momentum (under our approximations, this is the only
non-vanishing contribution to the relation between $v$ and
$G_\mu$). Splitting $K_r$ into SM and BSM parts, we find
\bea
\label{eq:KrSM}
16\,\pi^2\,K_r^{\smallSM} &=& \frac76\,N_c\,\frac{m_t^2}{v^2}~,\\[2mm]
\label{eq:KrBSM}
16\,\pi^2\,K_r^{\smallBSM} &=& -\frac{1}{v^2}\,\biggr[
  \frac{(\mHq-\MM)^2}{6\,\mHq}\,+\,\frac{(\mAq-\MM)^2}{6\,\mAq}
  \,+\,\frac{(\mHpq-\MM)^2}{3\,\mHpq}\nn\\[2mm]
  &&~~~~~~~~~~
-\,2\,\widetilde B_{22}(0,\mHq,\mHpq)
\,-\,2\,\widetilde B_{22}(0,\mAq,\mHpq)\biggr]~.
\eea
The divergent parts of the top-quark contributions to $\delta Z_{hh}$
and $\delta v^2/v^2$ cancel out against each other, leaving a residue
in eq.~(\ref{eq:KrSM}) that is finite and independent of the
renormalization scale. In contrast, the BSM parts of $\delta Z_{hh}$
and $\delta v^2/v^2$ are separately finite and scale-independent.  The
terms in the first line on the r.h.s.~of eq.~(\ref{eq:KrBSM}) stem
from $\delta Z_{hh}$, while the terms in the second line, where
\beq
\widetilde B_{22}(0,m_1^2,m_2^2)~=~
\frac12\,\left(\frac{m_1^2+m_2^2}{4}-\frac{m_1^2\,m_2^2}{2\,(m_1^2-m_2^2)}
\,\ln\frac{m_1^2}{m_2^2}\right)~,
\eeq
stem from $\delta v^2/v^2$.
Finally, we isolate the BSM part of the product $K_r\,{\cal P}_h^{1\ell}$\,:
\beq
\label{eq:KronP2BSM}
{\cal P}_h^{2\ell,\smallBSM}
~~\supset~~ K_r^{\smallBSM}\,{\cal P}_h^{1\ell,\smallSM} ~+~
(K_r^{\smallSM}+K_r^{\smallBSM})\,{\cal P}_h^{1\ell,\smallBSM}~,
\eeq
where ${\cal P}_h^{1\ell,\smallSM}$ is given in eq.~(\ref{eq:P1lSM}),
and ${\cal P}_h^{1\ell,\smallBSM}$ is the first term on the r.h.s.~of
eq.~(\ref{eq:P1lBSM}). We note that the first term on the r.h.s.~of
eq.~(\ref{eq:KronP2BSM}) above is enhanced by the relatively large
numerical value of the SM part of the one-loop amplitude, i.e., ${\cal
  P}_h^{1\ell,\smallSM} = -47/9$.

\bigskip

To validate our implementation of the LET of
refs.~\cite{Shifman:1979eb, Kniehl:1995tn}, we checked that we can
obtain the same result by computing directly the two-loop BSM
contributions to the $h\gamma\gamma$ amplitude, under the same
approximations employed in the calculation of
$\Pi_{\gamma\gamma}^{2\ell,\,\smallBSM}(0)$ (namely, the alignment
limit, the gaugeless limit and the vanishing of the SM-like Higgs
mass). We remark that this calculation involves counterterm
contributions analogous to the ones in eq.~(\ref{eq:dPh1l}), stemming
from the renormalization of the $h\, H^+H^-$ vertex and of the
charged-Higgs mass, and to the ones in eq.~(\ref{eq:KronP2BSM}),
stemming from the WFR of the external Higgs field and from the
renormalization of $v$.


As a second, non-trivial check, we verified that the BSM contributions
to the $h\gamma\gamma$ amplitude are independent of the renormalization
scale $Q$ at the perturbative order considered in our calculation:
\beq
\frac{d}{d\ln Q^2} \left({\cal P}_h^{1\ell,\smallBSM} \,+\, 
{\cal P}_h^{2\ell,\smallBSM} \right)~=~ 0~.
\eeq
This {\color{black} follows from the scale independence of
  $\Gamma(h\rightarrow\gamma\gamma)$, and} requires that we combine the
explicit scale dependence of ${\cal P}_h^{2\ell,\smallBSM}$ with the
implicit scale dependence of the parameters entering ${\cal
  P}_h^{1\ell,\smallBSM}$.  Of these, $M^2_{H^\pm}$ is defined as the
squared pole mass of the charged Higgs boson and is thus
scale-independent, but $\beta$ and $m_{12}^2$, which enter ${\cal
  P}_h^{1\ell,\smallBSM}$ in the combination $\MM = m_{12}^2/(s_\beta
c_\beta)$\,, are defined as $\msbar$-renormalized parameters. The
one-loop renormalization-group equation (RGE) for $\MM$
reads\,\footnote{We took the one-loop RGEs for the THDM parameters
  $\beta$ and $m_{12}^2$ from the code {\tt SARAH}~\cite{Staub:2008uz,
    Staub:2009bi, Staub:2010jh, Staub:2012pb, Staub:2013tta}. Formulas
  for these RGEs can also be found in ref.~\cite{BhupalDev:2014bir},
  but the coefficient of $y_t^2$ in the RGE for $m_{12}^2$ appears to
  be incorrect there.}
\beq
\label{eq:RGE}
16\pi^2\,\frac{d\,\MM}{d\ln Q^2} ~~=~~
\frac12\,N_c\,y_t^2\,c_{2\beta}\,\MM~+\,
\left(\lambda_3\,+\,2\,\lambda_4\,+\,3\,\lambda_5\,+\,\frac12\,N_c\,y_t^2\,
\right) \MM~,
\eeq
where we neglected the EW gauge couplings and the Yukawa couplings
other than $y_t$. The term proportional to $c_{2\beta}$ in
eq.~(\ref{eq:RGE}) stems from the RGE for $\beta$, and the rest stems
from the RGE for $m_{12}^2$.

{\color{black} Finally, we remark that our result for ${\cal
    P}_h^{2\ell,\smallBSM}$ does not vanish in the limit in which
  $\MM$ is pushed to infinity while the quartic Higgs couplings are
  kept fixed. The lack of decoupling behavior is due to our choice of
  an $\msbar$ definition for the parameter $\MM$ entering ${\cal
    P}_h^{1\ell,\smallBSM}$. The same issue was encountered in the
  calculation of the Higgs self-couplings of
  refs.~\cite{Braathen:2019pxr, Braathen:2019zoh}, where it was
  proposed that the non-decoupling terms be absorbed in a redefinition
  of the mass parameter.  Following that approach, we can define
  $(\MM)^{\rm dec} \,=\, (\MM)^{\msbar} \,+\, \delta \MM$, and we
  find:
  \beq
  \label{eq:deltaM2}
  \delta \MM ~=~ -\frac{\MM}{16\pi^2}\,\left[
    \left(\lambda_3\,+\,2\,\lambda_4\,+\,3\,\lambda_5\right)\,
    \left(1-\ln\frac{\MM}{Q^2}\right) \,+\,
    N_c\,y_t^2\,c_\beta^2\,\left(2-\ln\frac{\MM}{Q^2}\right)\,\right]\,.
  \eeq
  The combination of eqs.~(\ref{eq:RGE}) and (\ref{eq:deltaM2}) shows
  that $(\MM)^{\rm dec}$ is a scale-independent parameter.  If ${\cal
    P}_h^{1\ell,\smallBSM}$ is expressed in terms of $(\MM)^{\rm
    dec}$, ${\cal P}_h^{2\ell,\smallBSM}$ vanishes for
  $\MM\rightarrow\infty$, and in turn does not depend explicitly on
  the renormalization scale.  }

\section{Numerical impact of the two-loop BSM contributions to
  $\Gamma[h\rightarrow \gamma \gamma]$}
\label{sec:numerics}

We now illustrate the numerical impact of the newly-computed two-loop
corrections on the prediction for $\Gamma[h\rightarrow \gamma \gamma]$
in the aligned THDM. A comprehensive analysis of the parameter space
of the model along the lines of refs.~\cite{Arco:2020ucn,
  Arco:2022xum, Arco:2022jrt}, taking into account all of the
theoretical and experimental constraints, is well beyond the scope of
this paper. We will instead focus on two benchmark points introduced
in ref.~\cite{Bahl:2022xzi} to accommodate the recent CDF measurement
of the $W$ mass~\cite{CDF:2022hxs}, and discuss at the qualitative
level how the inclusion of the two-loop BSM contributions to
$\Gamma[h\rightarrow \gamma \gamma]$ can affect scenarios in which the
one-loop prediction is already in some tension with the experimental
value.

\bigskip

The two benchmark points introduced in ref.~\cite{Bahl:2022xzi} are
defined in terms of the three BSM Higgs masses, $m_H$, $m_A$ and
$m_{H^\pm}$, plus $\tan\beta$ and $\MM \equiv m_{12}^2/(s_\beta
c_\beta)$, while the angle $\alpha$ is fixed by the alignment
condition to $\beta-\pi/2$. The numerical values of the parameters
are\,\footnote{We rounded up the values of the parameters given in
table~I of ref.~\cite{Bahl:2022xzi}, but we checked that our results
remain essentially the same if we use the original values.}
\beq
\label{eq:pointA}
    {\rm Point~A\!:}~~~
    m_H\,=\,850~{\rm GeV},~~~
    m_A\,=\,930~{\rm GeV},~~~
    m_{H^\pm}\,=\,810~{\rm GeV},~~~
    \MM=(670~{\rm GeV})^2,~~
    \tan\beta\,=1.2\,,
\eeq
\beq
\label{eq:pointB}
    {\rm Point~B\!:}~~~
    m_H\,=\,350~{\rm GeV},~~~
    m_A\,=\,750~{\rm GeV},~~~
    m_{H^\pm}\,=\,760~{\rm GeV},~~~
    \MM=(235~{\rm GeV})^2,~~
    \tan\beta\,= 1.25\,.
\eeq
According to ref.~\cite{Bahl:2022xzi}, both of these points satisfy
the theoretical constraints of vacuum
stability~\cite{Barroso:2013awa}, boundedness from below of the Higgs
potential~\cite{Branco:2011iw}, and NLO perturbative
unitarity~\cite{Grinstein:2015rtl, Cacchio:2016qyh}. In addition, the
compatibility of the properties of the SM-like scalar $h$ with the
experimental measurements was checked with the code {\tt
  HiggsSignals}~\cite{Bechtle:2013xfa, Bechtle:2020uwn}, and the
constraints from direct searches of BSM Higgs bosons were checked with
the code {\tt HiggsBounds}~\cite{Bechtle:2008jh, Bechtle:2011sb,
  Bechtle:2013wla, Bechtle:2020pkv, Bahl:2021yhk}. Finally,
$b$-physics constraints were checked following
ref.~\cite{Haller:2018nnx}.

Large BSM contributions are required in order to yield a prediction
for $M_W$ of about $80.43$~GeV, compatible with the recent CDF
measurement~\cite{CDF:2022hxs} and $7\sigma$ away from the SM
prediction.  In aligned THDM scenarios, such contributions can stem
from large values of the quartic Higgs couplings. Indeed, in point~A
the couplings, extracted from the tree-level relations between Higgs
masses and Lagrangian parameters, include $\lambda_1 \approx 7$ and
$\lambda_5 \approx -7$, and in point~B they include $\lambda_3 \approx
16,~\lambda_4 \approx -9$ and $\lambda_5 \approx -8$. Crucially, the
authors of ref.~\cite{Bahl:2022xzi} point out that, in these
scenarios, the prediction for $M_W$ receives a significant shift from
the two-loop BSM contributions, which were obtained from
refs.~\cite{Hessenberger:2016atw, Hessenberger:2022tcx}.

To estimate the impact of such large couplings on the prediction for
$\Gamma[h\rightarrow \gamma \gamma]$, we define a {\color{black}
  simplified} signal-strength parameter
\beq
\label{eq:defmu}
\mu_{\gamma\gamma}~\equiv~\left|
\frac{{\cal P}_h^{\smallSM}\,+\,{\cal P}_h^{\smallBSM}}
   {{\cal P}_h^{\smallSM}}\right|^2~,
\eeq   
and we refer to $\mu_{\gamma\gamma}^{1\ell}$ when the BSM
contributions to the $h\gamma\gamma$ amplitude contain only the
one-loop part, i.e., when ${\cal P}_h^{\smallBSM} \,=\, {\cal
  P}_h^{1\ell,\,\smallBSM}$, and to $\mu_{\gamma\gamma}^{2\ell}$ when
they include also the newly-computed two-loop part, i.e., when ${\cal
  P}_h^{\smallBSM} \,=\, {\cal P}_h^{1\ell,\,\smallBSM} + {\cal
  P}_h^{2\ell,\,\smallBSM}$. Since we are only interested in a
qualitative discussion of the impact of the two-loop BSM
contributions, we do not include in ${\cal P}_h^{\smallSM}$ the full
two-loop result for the SM amplitude from refs.~\cite{Zheng:1990qa,
  Djouadi:1990aj, Dawson:1992cy, Melnikov:1993tj, Djouadi:1993ji,
  Inoue:1994jq, Fleischer:2004vb, Liao:1996td, Djouadi:1997rj,
  Fugel:2004ug, Degrassi:2005mc, Aglietti:2004nj, Aglietti:2004ki,
  Passarino:2007fp, Actis:2008ts}, but we approximate it with the
one-loop result in the limit of vanishing Higgs mass\,\footnote{Since
  the two-loop BSM amplitude is in turn computed in the limit of
  vanishing Higgs mass, it is even possible that this choice provides
  a better estimate of its effect on the signal-strength
  parameter. See, e.g., eq.~(\ref{eq:KronP2BSM}), where a numerically
  important contribution to ${\cal P}_h^{2\ell,\,\smallBSM}$ is in
  fact proportional to ${\cal P}_h^{1\ell,\,\smallSM}$\,.} from
eq.~(\ref{eq:P1lSM}), i.e., ${\cal P}_h^{\smallSM} \,=\, {\cal
  P}_h^{1\ell,\,\smallSM}\,=\,-47/9$. Note that an implicit assumption
in eq.~(\ref{eq:defmu}) is that the large quartic Higgs couplings do
not significantly affect the {\color{black} ratio of production cross
  section over total decay width of the SM-like Higgs boson. Indeed,
  the dominant next-to-leading order contributions involving those
  couplings affect the main production and decay channels through a
  common multiplicative factor $K_r^{\smallBSM}$, see
  eq.~(\ref{eq:KrBSM}), which cancels in the ratio.}

When the BSM contributions are computed at the one-loop level, we find
$\mu_{\gamma\gamma}^{1\ell} \approx 0.96$ in point~A and
$\mu_{\gamma\gamma}^{1\ell} \approx 0.89$ in point~B. Considering that
the LHC average of the signal strength for $h\longrightarrow
\gamma\gamma$ is currently $\mu_{\gamma\gamma}^{\rm exp}\,=\,1.10 \pm
0.07$~\cite{ParticleDataGroup:2022pth}, there appears to be some
tension at least with the prediction in point~B. This said, a
more-sophisticated determination of the {\color{black} signal strength}
could alter the picture somewhat, and even a scenario where the
prediction for $\mu_{\gamma\gamma}$ is more than $2\sigma$ away from
the measured value would not necessarily be ruled out in a global
analysis that takes into account a number of other physical
observables.

The inclusion of the two-loop BSM contributions requires that we
specify the renormalization scheme of the parameters entering the
one-loop BSM contributions, namely the mass of the charged Higgs boson
and the parameter $\MM$. We identify the former with the pole mass
$M_{H^{\pm}}$, and we interpret the latter as an $\msbar$-renormalized
parameter expressed at some scale $Q^2$. When we fix $\MM(Q^2)$ to the
numerical values given in eqs.~(\ref{eq:pointA}) and
(\ref{eq:pointB}), different choices for $Q^2$ correspond to different
points in the THDM parameter space. If, for example, we assume that
the values in eqs.~(\ref{eq:pointA}) and (\ref{eq:pointB}) correspond
to $\MM(\MM)$, we obtain $\mu_{\gamma\gamma}^{2\ell} \approx 0.95$ in
point~A and $\mu_{\gamma\gamma}^{2\ell} \approx 0.85$ in
point~B. While in point~A the impact of the two-loop BSM contributions
happens to be small, in point~B it might be large enough to turn
a scenario that was marginally allowed into an excluded one (a global
analysis that goes beyond the scope of this paper would be necessary
to reach a definite conclusion on this point). If we instead assume
that the values in eqs.~(\ref{eq:pointA}) and (\ref{eq:pointB})
correspond to $\MM(m_h^2)$, we obtain $\mu_{\gamma\gamma}^{2\ell}
\approx 0.92$ in point~A and $\mu_{\gamma\gamma}^{2\ell} \approx 0.84$
in point~B. Obviously, the effect of this change of scale is more
significant in point~A, where $\MM$ is farther away from $m_h^2$ than
in point~B.
{\color{black} Finally, if we assume that the values in
  eqs.~(\ref{eq:pointA}) and (\ref{eq:pointB}) correspond to the
  scale-independent parameter $(\MM)^{\rm dec}$ defined by
  eq.~(\ref{eq:deltaM2}) we find $\mu_{\gamma\gamma}^{2\ell} \approx
  0.94$ in point~A and $\mu_{\gamma\gamma}^{2\ell} \approx 0.84$ in
  point~B. In all}
cases it appears that the newly-computed ${\cal
  P}_h^{2\ell,\,\smallBSM}$ can amount to a significant fraction of
the total BSM contributions, and its inclusion may prove necessary to
obtain an accurate prediction for the $h\longrightarrow \gamma\gamma$
signal strength. We also remark that ${\cal P}_h^{2\ell,\,\smallBSM}$
is largely dominated by the contributions controlled by the quartic
Higgs couplings. Indeed, the results quoted above for
$\mu_{\gamma\gamma}^{2\ell}$ would hardly change if we were to neglect
the contributions to ${\cal P}_h^{2\ell,\,\smallBSM}$ controlled by
the top Yukawa coupling $y_t$.

\vfill
\newpage

The prediction for $\Gamma[h\rightarrow \gamma \gamma]$ has been used
in ref.~\cite{Arco:2022jrt} to study the constraints on the parameter
$\MM$ in regions of the aligned THDM that are allowed by all of the
other constraints, both theoretical and experimental. Once again, it
is legitimate to wonder how the inclusion of the two-loop BSM
contributions might alter the results of such a study. We will not
attempt here to repeat the extensive parameter scans of
ref.~\cite{Arco:2022jrt}, but we will just consider two scenarios
inspired by the benchmark points introduced above.

\begin{figure}[t]
\begin{center}
  \vspace*{-1.2cm}
  \includegraphics[width=8.3cm]{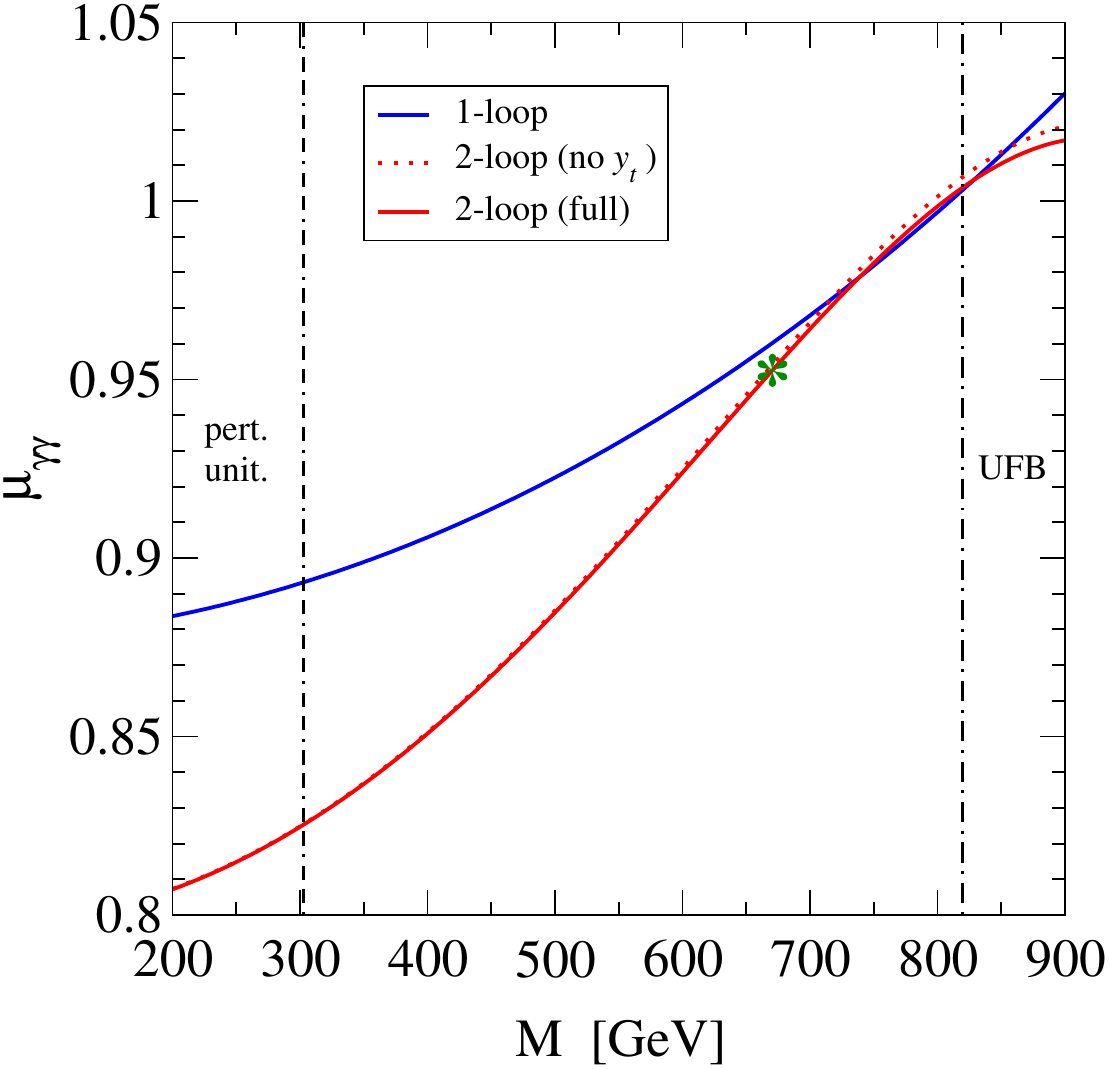}~~~
  \includegraphics[width=8.3cm]{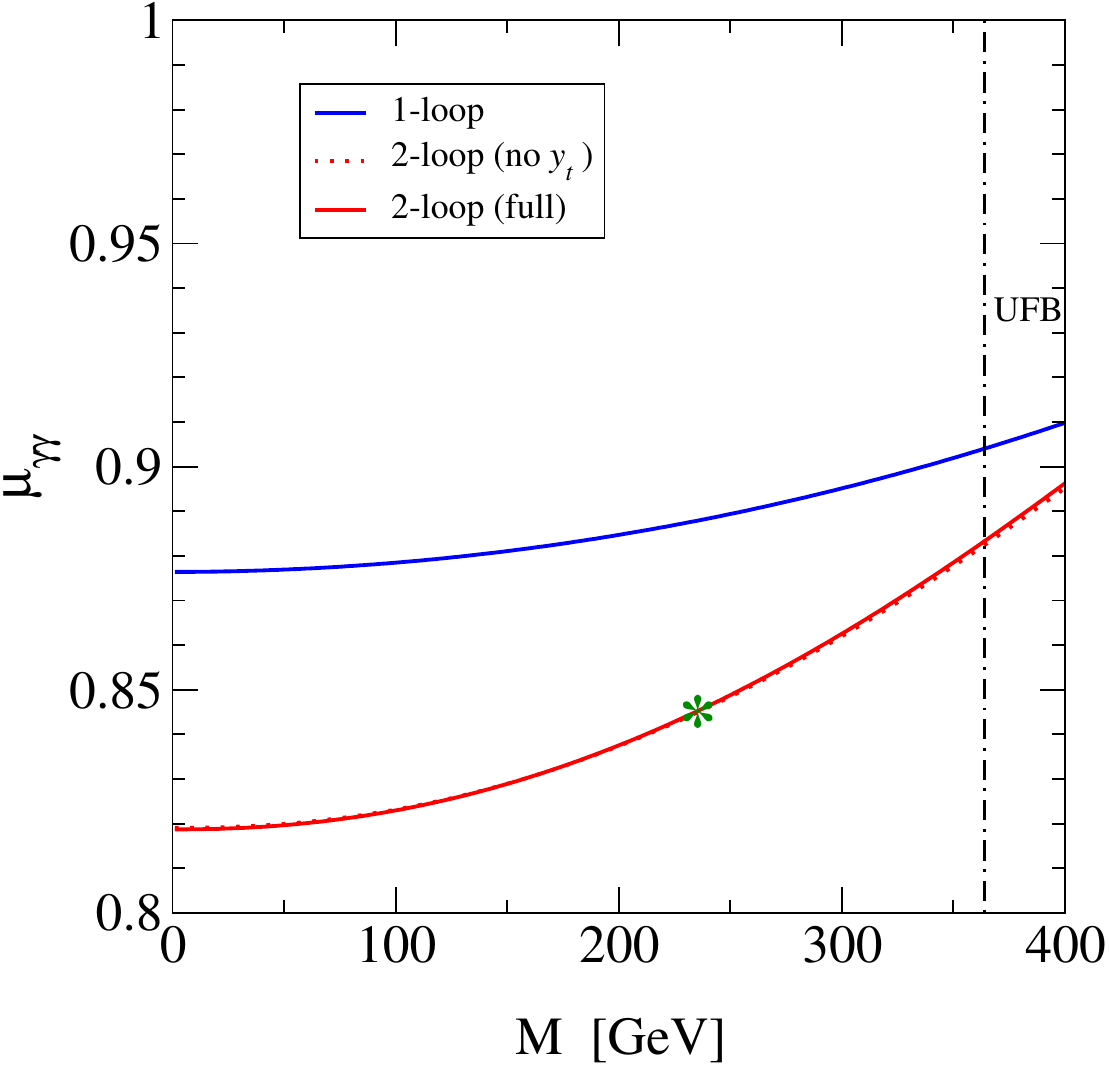}
  \caption{\em Signal strength for $h\longrightarrow \gamma\gamma$ as
    a function of $\sqrt{\MM}$. The remaining THDM parameters are
    fixed as in eq.~(\ref{eq:pointA}) (left plot) or as in
    eq.~(\ref{eq:pointB}) (right plot). The meaning of the different
    lines is explained in the text.}
  \label{figure}
\vspace*{-5mm}
\end{center}
\end{figure}

In figure~\ref{figure} we show the prediction for the signal strength
$\mu_{\gamma\gamma}$ as a function of the parameter $M~\,\equiv\,
\sqrt{\MM(\MM)}$, where the remaining THDM parameters are fixed as in
eq.~(\ref{eq:pointA}) for point~A (left plot) or as in
eq.~(\ref{eq:pointB}) for point~B (right plot). We do not consider any
experimental constraint on the parameter space, although most of them
presumably carry over from the analysis of ref.~\cite{Bahl:2022xzi}
since the BSM Higgs masses and $\tan\beta$ are the same as in the two
points introduced there. However, we do implement the theoretical
constraints from (tree-level) perturbative unitarity, boundedness from
below, and vacuum stability, following eqs.~(18)--(28) of
ref.~\cite{Arco:2020ucn}. In the region marked as ``UFB'' on the right
of the dot-dashed line in each plot, the quartic Higgs couplings
extracted from the tree-level relations between Higgs masses and
Lagrangian parameters violate at least one of the conditions for
boundedness from below. In the region on the left of the
dot-double-dashed line in the left plot, the couplings violate at
least one of the conditions for perturbative unitarity of the
scattering matrix. In the right plot the unitarity conditions are not
violated until $\MM \lesssim -(500~{\rm GeV})^2$, but negative values
  of $\MM$ violate the condition that ensures that the minimum
  of the potential is a global one.

Coming to the predictions for the signal strength, the solid blue line
in each plot represents $\mu_{\gamma\gamma}^{1\ell}$, the solid red
line represents our full computation of $\mu_{\gamma\gamma}^{2\ell}$,
whereas the dotted red line represents a computation of
$\mu_{\gamma\gamma}^{2\ell}$ in which the two-loop BSM contributions
controlled by the top Yukawa coupling have been omitted. The green
asterisk centered on the solid red line in the left and right plot
marks the value of $M$ that corresponds to the point of
eqs.~(\ref{eq:pointA}) and (\ref{eq:pointB}), respectively. Since we
adopt the limit of vanishing mass for the SM-like Higgs boson, the
dependence of the one-loop result on $M$ is given simply by
$\mu_{\gamma\gamma}^{1\ell} ~\approx~ 0.876 \,+\, 0.12\,x \,+\,
0.004\, x^2$, where $x= \MM/M^2_{H^\pm}$. Based on this naive estimate
of the signal strength,\footnote{Restoring the dependence of
  $\mu_{\gamma\gamma}^{1\ell}$ on $m_h^2$ would induce a shift of up
  to about $0.02$ in the lines of figure~\ref{figure}. However, this
  would not qualitatively alter our conclusions on the relevance of
  the two-loop BSM contributions.}  values of $M$ below about
$670$~GeV for the left plot and $630$~GeV for the right plot yield a
prediction for $\mu_{\gamma\gamma}^{1\ell}$ that, in these scenarios,
is more than $2\sigma$ away from the measured value.

The comparison between the blue lines for $\mu_{\gamma\gamma}^{1\ell}$
and the red lines for $\mu_{\gamma\gamma}^{2\ell}$ shows that the
two-loop BSM contributions can be significant, especially at lower
values of $M$. Being generally negative, they can exacerbate the
tension between measurement and theory prediction. Finally, the
comparison between the solid and dotted red lines shows that ${\cal
  P}_h^{2\ell,\,\smallBSM}$ is largely dominated by the contributions
controlled by the quartic Higgs couplings, except for a region at
large $M$ in the left plot where those contributions happen to be
rather small and are comparable in size with the contributions
controlled by the top Yukawa coupling. Once again, figure~\ref{figure}
shows that in the aligned THDM there are scenarios in which a precise
determination of $\Gamma[h\rightarrow \gamma \gamma]$ requires the
inclusion of the two-loop BSM contributions.

\section{Conclusions}
\label{sec:conclusions}

The requirement that an extension of the SM accommodate a scalar with
properties compatible with those observed at the LHC constrains the
parameter space of the BSM model even before the direct observation of
any new particles. In models such as the THDM, where some of the
couplings in the Lagrangian can take remarkably large values and still
be allowed by all theoretical constraints, precise predictions for the
properties of the SM-like Higgs boson $h$ and for other EW observables
may require that the contributions involving those couplings be
accounted for beyond the LO. Following the example of earlier two-loop
calculations of the $\rho$ parameter~\cite{Hessenberger:2016atw,
  Hessenberger:2022tcx}, of the scalar mass
matrices~\cite{Braathen:2017izn}, and of the trilinear self-coupling
of the SM-like Higgs boson~\cite{Braathen:2019pxr, Braathen:2019zoh},
in this work we computed the two-loop BSM contributions to
$\Gamma[h\rightarrow \gamma \gamma]$ in the aligned (and
CP-conserving) THDM. In line with the earlier calculations, we adopted
the simplifying assumptions of vanishing EW gauge couplings (the
so-called ``gaugeless limit'') and vanishing mass of the SM-like Higgs
boson. The latter assumption allowed us to exploit a LET that connects
the $h\gamma\gamma$ amplitude to the derivative of the photon
self-energy w.r.t.~the Higgs vev. In addition, the alignment and
gaugeless limits allowed us to adopt a simplified approach to the
renormalization of the mixing in the scalar sector, bypassing the
complications related to the possible gauge dependence of the mixing
angles that were discussed in refs.~\cite{Krause:2016oke,
  Altenkamp:2017ldc, Kanemura:2017wtm}. We provided explicit analytic
formulas for the two-loop BSM contributions to the photon
self-energy. The corresponding formulas for the $h\gamma\gamma$
amplitude can be obtained straightforwardly by exploiting the chain
rule for the derivative w.r.t.~$v$, and we make them available on
request in electronic form.

After describing our calculation, we briefly discussed the numerical
impact of the newly-computed two-loop BSM contributions. We chose not
to embark in an extensive analysis of the THDM parameter space, but
rather focus on two benchmark points introduced in
ref.~\cite{Bahl:2022xzi}, where large values of the quartic Higgs
couplings lead to a prediction for the $W$ mass compatible with the recent
CDF measurement~\cite{CDF:2022hxs}, and $7\sigma$ away from the SM
prediction. As expected, such large couplings can also lead to a
deviation from the SM prediction for the $h\longrightarrow \gamma\gamma$
decay width. We defined a simplified signal-strength parameter
$\mu_{\gamma\gamma}$, and showed how the inclusion of the two-loop BSM
contributions can exacerbate the tension between the measured value of
the signal strength, which is in fact slightly above the SM
prediction, and the prediction of the THDM in the considered benchmark
points, which is somewhat below the SM prediction. We also discussed
how the prediction for $\mu_{\gamma\gamma}$ depends on the value and,
beyond the LO, the definition of the parameter $\MM$. In summary, we
showed that in the aligned THDM there are scenarios in which the
inclusion of the two-loop BSM contributions is required for a precise
determination of $\Gamma[h\rightarrow \gamma \gamma]$.

Of course, our discussion of the numerical impact of our results
should be viewed as merely qualitative. A detailed study of the
constraints on the aligned THDM arising from $\Gamma[h\rightarrow
  \gamma \gamma]$ would require that we combine the BSM contributions
to the $h\gamma\gamma$ amplitude with the most complete determination
of the SM contributions, {\color{black} that we account for the BSM
  corrections to all production and decay processes,} that we perform
extensive scans over the parameter space of the model, and that we
take into account all of the remaining theoretical and experimental
constraints. We leave such analysis for future work, hoping that the
results presented in this paper will help the collective effort to use
the properties of the Higgs boson as a probe of what lies beyond the
SM.

\vfill

\section*{Acknowledgments}

We thank H.~Haber for discussions that sparked our interest in this
topic, and M.~Goodsell for useful communications about the results of
{\tt SARAH} for the RGEs of the THDM. {\color{black} We also thank
  J.~Braathen for useful communications about
  ref.~\cite{Aiko:2023nqj}, which was posted on the arXiv shortly
  after our paper and deals with the $h\longrightarrow\gamma\gamma$
  decay in a different variant of the THDM.}  The work of G.~D.~is
partially supported by the Italian Ministry of Research (MUR) under
the grant PRIN 20172LNEEZ.

\vfill
\newpage

\section*{Appendix A: One-loop self-energies and tadpoles of the Higgs bosons}
\setcounter{equation}{0}
\renewcommand{\theequation}{A\arabic{equation}}

In this appendix we list explicit formulas for the one-loop
self-energies and tadpoles of the Higgs bosons that are relevant to
our calculation. They were obtained by adapting to the THDM the
general formulas given in refs.~\cite{Braathen:2016cqe,
  Braathen:2018htl}, under the limits of alignment, vanishing EW gauge
couplings and vanishing SM-like Higgs mass.
\bea
16\pi^2\, \Pi_{H^+H^-}(\mHpq) &=&
-\frac{1}{v^2}\,\biggr[
  (\mHpq-\mHq)^2\,B_0(\mHpq,\mHq,0)
  \,+\,(\mHpq-\mAq)^2\,B_0(\mHpq,\mAq,0)\nn\\
  &&~~~~~~~+4\,(\mHpq-\MM)^2\,B_0(\mHpq,\mHpq,0)\nn\\
  &&~~~~~~~-2\,(\mHq-\MM)\cot^22\beta\,
  \biggr(A_0(\mHq)+A_0(\mAq)+4\,A_0(\mHpq)\nn\\
  &&~~~~~~~~~~~~~~~~~~~~~~~~~~~~~~~~~~~~~~~~
  -2\,(\mHq-\MM)\,B_0(\mHpq,\mHpq,\mHq)
  \biggr)\,\biggr]\nn\\
&-& \frac{2\,m_t^2}{v^2}\,N_c\,\cot^2\beta \, G_0(\mHpq,0,m_t^2)~,
\label{eq:PiHpHm}
\eea
\bea
16\pi^2\, \Pi_{hh}(0) &=&
\frac{1}{v^2}\,\biggr[
  (\mHq-\MM)\,\biggr( A_0(\mHq) - 2\,(\mHq-\MM)\,B_0(0,\mHq,\mHq)\biggr)\nn\\
  &&~~~~~+\,
  (\mAq-\MM)\,\biggr( A_0(\mAq) - 2\,(\mAq-\MM)\,B_0(0,\mAq,\mAq)\biggr)\nn\\
  &&~~~~~+\,
  2\,(\mHpq-\MM)\,\biggr( A_0(\mHpq)
  - 2\,(\mHpq-\MM)\,B_0(0,\mHpq,\mHpq)\biggr)\biggr]\nn\\
&+& \frac{2\, m_t^2}{v^2}\,N_c\, 
\biggr( 2\,m_t^2\,B_0(0,m_t^2,m_t^2) - G_0(0,m_t^2,m_t^2)\biggr)~,
\eea  
\bea
16\pi^2\, \Pi_{hH}(0) &=&
\frac{\MM-\mHq}{v^2}\,\cot2\beta\,\biggr[
  3\,\biggr( A_0(\mHq) - 2\,(\mHq-\MM)\,B_0(0,\mHq,\mHq)\biggr)\nn\\
  &&~~~~~~~~~~~~~~~~~~~~~~~~+\,
  A_0(\mAq) - 2\,(\mAq-\MM)\,B_0(0,\mAq,\mAq)\nn\\
  &&~~~~~~~~~~~~~~~~~~~~~~~\,+\,
  2\,\biggr( A_0(\mHpq)
  - 2\,(\mHpq-\MM)\,B_0(0,\mHpq,\mHpq)\biggr)\biggr]\nn\\
&-& \frac{2\,m_t^2}{v^2}\,N_c\, \cot\beta\,
\biggr( 2\,m_t^2\,B_0(0,m_t^2,m_t^2) - G_0(0,m_t^2,m_t^2)\biggr)~,
\eea  
\bea
16\pi^2\, T_h &=&
\frac{1}{v}\,\biggr[(\mHq-\MM)\,A_0(\mHq)\,+\,(\mAq-\MM)\,A_0(\mAq)
  \,+\,2\,(\mHpq-\MM)\,A_0(\mHpq)\biggr]\nn\\
&-&\frac{4\,m_t^2}{v}\,N_c\,A_0(m_t^2)~,\\[2mm]
16\pi^2\, T_H &=&
\frac{\MM-\mHq}{v}\,\cot2\beta\,\biggr[
  3\,A_0(\mHq) \,+\, A_0(\mAq) \,+\, 2\, A_0(\mHpq)\biggr]\nn\\
&+&\frac{4\,m_t^2}{v}\,N_c\,\cot\beta\,A_0(m_t^2)
~,
\eea
where
\beq
G_0(p^2,m_1^2,m_2^2)~=~ (p^2-m_1^2-m_2^2)\,B_0(p^2,m_1^2,m_2^2)
+ A_0(m_1^2) + A_0(m_2^2)~,
\eeq
and
\bea
\label{eq:A0}
A_0(m^2) &=& m^2\,\left(\ln\frac{m^2}{Q^2}-1\right)~,\\[2mm]
B_0(0,m^2,m^2) &=& -\ln\frac{m^2}{Q^2}~,\\[2mm]
{\rm Re}\,B_0(p^2,0,m^2) \,=\, {\rm Re}\,B_0(p^2,m^2,0) &=&
2\,-\,\ln\frac{m^2}{Q^2}\,
-\,\left(1-\frac{m^2}{p^2}\right)\,\ln\left|1-\frac{p^2}{m^2}\right|.
\eea
For the function $B_0(\mHpq,\mHpq,\mHq)$ in eq.~(\ref{eq:PiHpHm}) we
used the code {\tt LoopTools}~\cite{Hahn:1998yk}.  We checked that our
result for the charged-Higgs self-energy agrees with the corresponding
result in ref.~\cite{Braathen:2019zoh}.

\section*{Appendix B: Two-loop self-energy of the photon}
\setcounter{equation}{0}
\renewcommand{\theequation}{B\arabic{equation}}

Under the assumptions of alignment, vanishing EW gauge couplings, and
vanishing SM-like Higgs mass, the BSM part of the two-loop self-energy
of the photon reads

\bea
\Pi_{\gamma\gamma}^{2\ell,\,\smallBSM}(0)&=&
\frac{\alpha}{48\pi^3v^2}\,\biggr[
  ~(\mHpq-\MM)^2\,F_h(\mHpq)\nn\\[2mm]
  &&~~~~~~~~~~~+\,\frac{(\mHq-\mHpq)^2}{4}\,F_G(\mHq,\mHpq)\nn\\[2mm]
  &&~~~~~~~~~~~+\,\frac{(\mAq-\mHpq)^2}{4}\,F_G(\mAq,\mHpq) \nn\\[2mm]
  &&~~~~~~~~~~~+\,(\mHq-\MM)^2\,\cot^22\beta\,F_H(\mHq,\mHpq)\nn\\[2mm]
  &&~~~~~~~~~~~+\,\frac{\mHq-\MM}{2\,\mHpq}\,\cot^22\beta\,
  \biggr(A_0(\mHq)\,+\,A_0(\mAq)\,+4\,\,A_0(\mHpq)\biggr)\nn\\[2mm]
  &&~~~~~~~~~~~+ m_t^2\,\cot^2\beta\,N_c\,
  \biggr(f_H(m_t^2/\mHq)\,+\,f_A(m_t^2/\mAq)\,+\,f_{H^\pm}(m_t^2/\mHpq)\nn\\[2mm]
  &&~~~~~~~~~~~~~~~~~~~~~~~~~~~~~~~~~~~
  -\frac{A_0(m_t^2)}{\mHpq} \,-\,\frac{43}{72}
  \,-\,\frac13\,\ln\frac{m_t^2}{Q^2}\,\biggr)\biggr]~,
\label{eq:Pigaga2l}
\eea
where we also assume that the one-loop
{\color{black} part of the photon self-energy, see eq.~(\ref{eq:Pigaga1l}), is
  expressed in terms of $\msbar$-renormalized masses, and in particular
  the charged-Higgs contribution is expressed in terms of the
  parameter}
$\widehat m^2_{H^\pm}$ defined in eq.~(\ref{eq:mHphat}). If the
one-loop part was instead expressed in terms of the {\color{black}
  parameter} $\widetilde m^2_{H^\pm}$ defined in
eq.~(\ref{eq:mHptil}), $\Pi_{\gamma\gamma}^{2\ell,\,\smallBSM}(0)$
would receive an additional contribution corresponding to the second
term within parentheses in eq.~(\ref{eq:shiftPigaga}).

\vfill
The function $A_0(m^2)$ entering eq.~(\ref{eq:Pigaga2l}) is defined in
eq.~(\ref{eq:A0}), and the remaining functions are
\beq
F_h(m^2)~=~\frac{1}{m^2}\left(\ln\frac{m^2}{Q^2}-\frac12\right)~,
\eeq

\beq
F_G(m_1^2,m_2^2)~=~\frac{1}{m_2^2}\left(\ln\frac{m_1^2}{Q^2}-1\right)
\,-\,\frac{2\,(4\,m_1^2+5\,m_2^2)}{(m_1^2-m_2^2)^2}
\,+\,\frac{m_2^2\,(17\,m_1^2+m_2^2)}{(m_1^2-m_2^2)^3}\,\ln\frac{m_1^2}{m_2^2}~,
\eeq  

\beq
F_H(m_1^2,m_2^2)~=~\frac{1}{m_2^2}\left(\ln\frac{m_1^2}{Q^2}-1\right)
\,+\,\frac{1}{m_1^2-4\,m_2^2}
\,+\,\frac{m_1^2-10\,m_2^2}{(m_1^2-4\,m_2^2)^2}\,\ln\frac{m_1^2}{m_2^2}
\,+\,\frac{6\,m_2^4}{m_1^2\,(m_1^2-4\,m_2^2)^2}
\,\phi\left(\frac{m_1^2}{4\,m_2^2}\right)~,
\label{eq:FH}
\eeq  

\beq
f_H(x)~=~ \frac{-2}{3\,(1-4\,x)}\,\left[(1+2\,x)\ln x \,
  +\,2\,x\,(1-x)\,\phi(\frac1{4x})\right]~,
\eeq

\beq
f_A(x)~=~ \frac{2}{9\,(1-4\,x)^2}\,\left[1\,-\,4\,x\,+\,(5-14\,x)\ln x \,
  +\,6\,x\,(1-3\,x)\,\phi(\frac1{4x})\right]~,
\label{eq:fA}
\eeq

\beq
f_{H^{\pm}}(x)~=~ -\frac{\ln x}{9\,(1-x)}~.
\eeq

The function $\phi(z)$ entering eqs.~(\ref{eq:FH})--(\ref{eq:fA}) is
defined as
\beq
\phi(z) = \left\{
\begin{tabular}{ll}
  $4\, \sqrt{\frac{z}{1-z}} ~{\rm Cl}_2 ( 2 \arcsin \sqrt z )$ \, , &  
  $(0 < z < 1)$ \, , \\ \\
  ${ \frac1{\lambda} \left[ - 4 \,{\rm Li_2} (\frac{1-\lambda}2) +
    2\, \ln^2 (\frac{1-\lambda}2) - \ln^2 (4z) +\pi^2/3 \right] }$ \, ,
  & ~~~$(z \ge 1)$ \,,
\end{tabular}
\label{eq:phi}
\right.
\eeq
where ${\rm Cl}_2(z)= {\rm Im} \,{\rm Li_2} (e^{iz})$ is the Clausen
function, and $\lambda = \sqrt{1 - (1 / z)}$. A recursive relation for
the derivative of $\phi(z)$,
\beq
\frac{d}{dz}\,\phi(z)
~=~ \frac{2}{z-1}\,\left(\ln 4z-\frac{\phi(z)}{4z}\right)~,
\eeq
proves very useful in obtaining compact results for the derivatives of
$\Pi_{\gamma\gamma}^{2\ell,\,\smallBSM}(0)$ w.r.t.~the masses of the
top quark and of the BSM Higgs bosons, see eq.~(\ref{eq:catenone}).

\vfill
\newpage

\bibliographystyle{utphys}
\bibliography{THDM}

\providecommand{\href}[2]{#2}\begingroup\raggedright\begin{thebibliography}{10}

\bibitem{CMS:2012qbp}
{\bf CMS} Collaboration, S.~Chatrchyan {\em et al.}, {\em {Observation of a New
  Boson at a Mass of 125 GeV with the CMS Experiment at the LHC}}.
  \href{http://dx.doi.org/10.1016/j.physletb.2012.08.021}{Phys. Lett. B {\bf
  716} (2012)  30--61}, \href{http://arxiv.org/abs/1207.7235}{{\tt
  arXiv:1207.7235 [hep-ex]}}.

\bibitem{ATLAS:2012yve}
{\bf ATLAS} Collaboration, G.~Aad {\em et al.}, {\em {Observation of a new
  particle in the search for the Standard Model Higgs boson with the ATLAS
  detector at the LHC}}.
  \href{http://dx.doi.org/10.1016/j.physletb.2012.08.020}{Phys. Lett. B {\bf
  716} (2012)  1--29}, \href{http://arxiv.org/abs/1207.7214}{{\tt
  arXiv:1207.7214 [hep-ex]}}.

\bibitem{ATLAS:2015yey}
{\bf ATLAS, CMS} Collaboration, G.~Aad {\em et al.}, {\em {Combined Measurement
  of the Higgs Boson Mass in $pp$ Collisions at $\sqrt{s}=7$ and 8 TeV with the
  ATLAS and CMS Experiments}}.
  \href{http://dx.doi.org/10.1103/PhysRevLett.114.191803}{Phys. Rev. Lett. {\bf
  114} (2015)  191803}, \href{http://arxiv.org/abs/1503.07589}{{\tt
  arXiv:1503.07589 [hep-ex]}}.

\bibitem{ATLAS:2016neq}
{\bf ATLAS, CMS} Collaboration, G.~Aad {\em et al.}, {\em {Measurements of the
  Higgs boson production and decay rates and constraints on its couplings from
  a combined ATLAS and CMS analysis of the LHC pp collision data at $
  \sqrt{s}=7 $ and 8 TeV}}.
  \href{http://dx.doi.org/10.1007/JHEP08(2016)045}{JHEP {\bf 08} (2016)  045},
  \href{http://arxiv.org/abs/1606.02266}{{\tt arXiv:1606.02266 [hep-ex]}}.

\bibitem{Gunion:1989we}
J.~F. Gunion, H.~E. Haber, G.~L. Kane, and S.~Dawson, {\em {The Higgs Hunter's
  Guide}}. Front. Phys. {\bf 80} (2000)  .

\bibitem{Aoki:2009ha}
M.~Aoki, S.~Kanemura, K.~Tsumura, and K.~Yagyu, {\em {Models of Yukawa
  interaction in the two Higgs doublet model, and their collider
  phenomenology}}. \href{http://dx.doi.org/10.1103/PhysRevD.80.015017}{Phys.
  Rev. D {\bf 80} (2009)  015017}, \href{http://arxiv.org/abs/0902.4665}{{\tt
  arXiv:0902.4665 [hep-ph]}}.

\bibitem{Branco:2011iw}
G.~C. Branco, P.~M. Ferreira, L.~Lavoura, M.~N. Rebelo, M.~Sher, and J.~P.
  Silva, {\em {Theory and phenomenology of two-Higgs-doublet models}}.
  \href{http://dx.doi.org/10.1016/j.physrep.2012.02.002}{Phys. Rept. {\bf 516}
  (2012)  1--102}, \href{http://arxiv.org/abs/1106.0034}{{\tt arXiv:1106.0034
  [hep-ph]}}.

\bibitem{Gunion:2002zf}
J.~F. Gunion and H.~E. Haber, {\em {The CP conserving two Higgs doublet model:
  The Approach to the decoupling limit}}.
  \href{http://dx.doi.org/10.1103/PhysRevD.67.075019}{Phys. Rev. D {\bf 67}
  (2003)  075019}, \href{http://arxiv.org/abs/hep-ph/0207010}{{\tt
  arXiv:hep-ph/0207010}}.

\bibitem{Antoniadis:2006uj}
I.~Antoniadis, K.~Benakli, A.~Delgado, and M.~Quiros, {\em {A New gauge
  mediation theory}}. Adv. Stud. Theor. Phys. {\bf 2} (2008)  645--672,
  \href{http://arxiv.org/abs/hep-ph/0610265}{{\tt arXiv:hep-ph/0610265}}.

\bibitem{Carena:2013ooa}
M.~Carena, I.~Low, N.~R. Shah, and C.~E.~M. Wagner, {\em {Impersonating the
  Standard Model Higgs Boson: Alignment without Decoupling}}.
  \href{http://dx.doi.org/10.1007/JHEP04(2014)015}{JHEP {\bf 04} (2014)  015},
  \href{http://arxiv.org/abs/1310.2248}{{\tt arXiv:1310.2248 [hep-ph]}}.

\bibitem{Arco:2020ucn}
F.~Arco, S.~Heinemeyer, and M.~J. Herrero, {\em {Exploring sizable triple Higgs
  couplings in the 2HDM}}.
  \href{http://dx.doi.org/10.1140/epjc/s10052-020-8406-8}{Eur. Phys. J. C {\bf
  80} (2020) no.~9, 884}, \href{http://arxiv.org/abs/2005.10576}{{\tt
  arXiv:2005.10576 [hep-ph]}}.

\bibitem{Arco:2022xum}
F.~Arco, S.~Heinemeyer, and M.~J. Herrero, {\em {Triple Higgs couplings in the
  2HDM: the complete picture}}.
  \href{http://dx.doi.org/10.1140/epjc/s10052-022-10485-9}{Eur. Phys. J. C {\bf
  82} (2022) no.~6, 536}, \href{http://arxiv.org/abs/2203.12684}{{\tt
  arXiv:2203.12684 [hep-ph]}}.

\bibitem{Bahl:2022xzi}
H.~Bahl, J.~Braathen, and G.~Weiglein, {\em {New physics effects on the W-boson
  mass from a doublet extension of the SM Higgs sector}}.
  \href{http://dx.doi.org/10.1016/j.physletb.2022.137295}{Phys. Lett. B {\bf
  833} (2022)  137295}, \href{http://arxiv.org/abs/2204.05269}{{\tt
  arXiv:2204.05269 [hep-ph]}}.

\bibitem{CDF:2022hxs}
{\bf CDF} Collaboration, T.~Aaltonen {\em et al.}, {\em {High-precision
  measurement of the $W$ boson mass with the CDF II detector}}.
  \href{http://dx.doi.org/10.1126/science.abk1781}{Science {\bf 376} (2022)
  no.~6589, 170--176}.

\bibitem{Hessenberger:2016atw}
S.~Hessenberger and W.~Hollik, {\em {Two-loop corrections to the $\rho$
  parameter in Two-Higgs-Doublet Models}}.
  \href{http://dx.doi.org/10.1140/epjc/s10052-017-4734-8}{Eur. Phys. J. C {\bf
  77} (2017) no.~3, 178}, \href{http://arxiv.org/abs/1607.04610}{{\tt
  arXiv:1607.04610 [hep-ph]}}.

\bibitem{Hessenberger:2022tcx}
S.~Hessenberger and W.~Hollik, {\em {Two-loop improved predictions for $\mathbf
  {M_W}$ and $\mathbf {sin^2\theta _{eff}}$ in Two-Higgs-Doublet models}}.
  \href{http://dx.doi.org/10.1140/epjc/s10052-022-10933-6}{Eur. Phys. J. C {\bf
  82} (2022) no.~10, 970}, \href{http://arxiv.org/abs/2207.03845}{{\tt
  arXiv:2207.03845 [hep-ph]}}.

\bibitem{Braathen:2017izn}
J.~Braathen, M.~D. Goodsell, and F.~Staub, {\em {Supersymmetric and
  non-supersymmetric models without catastrophic Goldstone bosons}}.
  \href{http://dx.doi.org/10.1140/epjc/s10052-017-5303-x}{Eur. Phys. J. C {\bf
  77} (2017) no.~11, 757}, \href{http://arxiv.org/abs/1706.05372}{{\tt
  arXiv:1706.05372 [hep-ph]}}.

\bibitem{Braathen:2019pxr}
J.~Braathen and S.~Kanemura, {\em {On two-loop corrections to the Higgs
  trilinear coupling in models with extended scalar sectors}}.
  \href{http://dx.doi.org/10.1016/j.physletb.2019.07.021}{Phys. Lett. B {\bf
  796} (2019)  38--46}, \href{http://arxiv.org/abs/1903.05417}{{\tt
  arXiv:1903.05417 [hep-ph]}}.

\bibitem{Braathen:2019zoh}
J.~Braathen and S.~Kanemura, {\em {Leading two-loop corrections to the Higgs
  boson self-couplings in models with extended scalar sectors}}.
  \href{http://dx.doi.org/10.1140/epjc/s10052-020-7723-2}{Eur. Phys. J. C {\bf
  80} (2020) no.~3, 227}, \href{http://arxiv.org/abs/1911.11507}{{\tt
  arXiv:1911.11507 [hep-ph]}}.

\bibitem{ParticleDataGroup:2022pth}
{\bf Particle Data Group} Collaboration, R.~L. Workman {\em et al.}, {\em
  {Review of Particle Physics}}.
  \href{http://dx.doi.org/10.1093/ptep/ptac097}{PTEP {\bf 2022} (2022)
  083C01}.

\bibitem{Arco:2022jrt}
F.~Arco, S.~Heinemeyer, and M.~J. Herrero, {\em {Sensitivity and constraints to
  the 2HDM soft-breaking $Z_2$ parameter $m_{12}$}}.
  \href{http://dx.doi.org/10.1016/j.physletb.2022.137548}{Phys. Lett. B {\bf
  835} (2022)  137548}, \href{http://arxiv.org/abs/2207.13501}{{\tt
  arXiv:2207.13501 [hep-ph]}}.

\bibitem{Shifman:1979eb}
M.~A. Shifman, A.~I. Vainshtein, M.~B. Voloshin, and V.~I. Zakharov, {\em
  {Low-Energy Theorems for Higgs Boson Couplings to Photons}}. Sov. J. Nucl.
  Phys. {\bf 30} (1979)  711--716.

\bibitem{Kniehl:1995tn}
B.~A. Kniehl and M.~Spira, {\em {Low-energy theorems in Higgs physics}}.
  \href{http://dx.doi.org/10.1007/s002880050007}{Z. Phys. C {\bf 69} (1995)
  77--88}, \href{http://arxiv.org/abs/hep-ph/9505225}{{\tt
  arXiv:hep-ph/9505225}}.

\bibitem{Davidson:2005cw}
S.~Davidson and H.~E. Haber, {\em {Basis-independent methods for the
  two-Higgs-doublet model}}.
  \href{http://dx.doi.org/10.1103/PhysRevD.72.099902}{Phys. Rev. D {\bf 72}
  (2005)  035004}, \href{http://arxiv.org/abs/hep-ph/0504050}{{\tt
  arXiv:hep-ph/0504050}}. [Erratum: Phys.Rev.D 72, 099902 (2005)].

\bibitem{Kanemura:2004mg}
S.~Kanemura, Y.~Okada, E.~Senaha, and C.~P. Yuan, {\em {Higgs coupling
  constants as a probe of new physics}}.
  \href{http://dx.doi.org/10.1103/PhysRevD.70.115002}{Phys. Rev. D {\bf 70}
  (2004)  115002}, \href{http://arxiv.org/abs/hep-ph/0408364}{{\tt
  arXiv:hep-ph/0408364}}.

\bibitem{Krause:2016oke}
M.~Krause, R.~Lorenz, M.~Muhlleitner, R.~Santos, and H.~Ziesche, {\em
  {Gauge-independent Renormalization of the 2-Higgs-Doublet Model}}.
  \href{http://dx.doi.org/10.1007/JHEP09(2016)143}{JHEP {\bf 09} (2016)  143},
  \href{http://arxiv.org/abs/1605.04853}{{\tt arXiv:1605.04853 [hep-ph]}}.

\bibitem{Altenkamp:2017ldc}
L.~Altenkamp, S.~Dittmaier, and H.~Rzehak, {\em {Renormalization schemes for
  the Two-Higgs-Doublet Model and applications to h $\rightarrow WW/ZZ
  \rightarrow 4$ fermions}}.
  \href{http://dx.doi.org/10.1007/JHEP09(2017)134}{JHEP {\bf 09} (2017)  134},
  \href{http://arxiv.org/abs/1704.02645}{{\tt arXiv:1704.02645 [hep-ph]}}.

\bibitem{Kanemura:2017wtm}
S.~Kanemura, M.~Kikuchi, K.~Sakurai, and K.~Yagyu, {\em {Gauge invariant
  one-loop corrections to Higgs boson couplings in non-minimal Higgs models}}.
  \href{http://dx.doi.org/10.1103/PhysRevD.96.035014}{Phys. Rev. D {\bf 96}
  (2017) no.~3, 035014}, \href{http://arxiv.org/abs/1705.05399}{{\tt
  arXiv:1705.05399 [hep-ph]}}.

\bibitem{Zheng:1990qa}
H.-Q. Zheng and D.-D. Wu, {\em {First order QCD corrections to the decay of the
  Higgs boson into two photons}}.
  \href{http://dx.doi.org/10.1103/PhysRevD.42.3760}{Phys. Rev. D {\bf 42}
  (1990)  3760--3763}.

\bibitem{Djouadi:1990aj}
A.~Djouadi, M.~Spira, J.~J. van~der Bij, and P.~M. Zerwas, {\em {QCD
  corrections to gamma gamma decays of Higgs particles in the intermediate mass
  range}}. \href{http://dx.doi.org/10.1016/0370-2693(91)90879-U}{Phys. Lett. B
  {\bf 257} (1991)  187--190}.

\bibitem{Dawson:1992cy}
S.~Dawson and R.~P. Kauffman, {\em {QCD corrections to H ---\ensuremath{>}
  gamma gamma}}. \href{http://dx.doi.org/10.1103/PhysRevD.47.1264}{Phys. Rev. D
  {\bf 47} (1993)  1264--1267}.

\bibitem{Melnikov:1993tj}
K.~Melnikov and O.~I. Yakovlev, {\em {Higgs ---\ensuremath{>} two photon decay:
  QCD radiative correction}}.
  \href{http://dx.doi.org/10.1016/0370-2693(93)90507-E}{Phys. Lett. B {\bf 312}
  (1993)  179--183}, \href{http://arxiv.org/abs/hep-ph/9302281}{{\tt
  arXiv:hep-ph/9302281}}.

\bibitem{Djouadi:1993ji}
A.~Djouadi, M.~Spira, and P.~M. Zerwas, {\em {Two photon decay widths of Higgs
  particles}}. \href{http://dx.doi.org/10.1016/0370-2693(93)90564-X}{Phys.
  Lett. B {\bf 311} (1993)  255--260},
  \href{http://arxiv.org/abs/hep-ph/9305335}{{\tt arXiv:hep-ph/9305335}}.

\bibitem{Inoue:1994jq}
M.~Inoue, R.~Najima, T.~Oka, and J.~Saito, {\em {QCD corrections to two photon
  decay of the Higgs boson and its reverse process}}.
  \href{http://dx.doi.org/10.1142/S0217732394001003}{Mod. Phys. Lett. A {\bf 9}
  (1994)  1189--1194}.

\bibitem{Fleischer:2004vb}
J.~Fleischer, O.~V. Tarasov, and V.~O. Tarasov, {\em {Analytical result for the
  two loop QCD correction to the decay H ---\ensuremath{>} 2 gamma}}.
  \href{http://dx.doi.org/10.1016/j.physletb.2004.01.063}{Phys. Lett. B {\bf
  584} (2004)  294--297}, \href{http://arxiv.org/abs/hep-ph/0401090}{{\tt
  arXiv:hep-ph/0401090}}.

\bibitem{Liao:1996td}
Y.~Liao and X.-y. Li, {\em {O (alpha**2 G(F)m(t)**2) contributions to H
  ---\ensuremath{>} gamma gamma}}.
  \href{http://dx.doi.org/10.1016/S0370-2693(97)00089-0}{Phys. Lett. B {\bf
  396} (1997)  225--230}, \href{http://arxiv.org/abs/hep-ph/9605310}{{\tt
  arXiv:hep-ph/9605310}}.

\bibitem{Djouadi:1997rj}
A.~Djouadi, P.~Gambino, and B.~A. Kniehl, {\em {Two loop electroweak heavy
  fermion corrections to Higgs boson production and decay}}.
  \href{http://dx.doi.org/10.1016/S0550-3213(98)00147-3}{Nucl. Phys. B {\bf
  523} (1998)  17--39}, \href{http://arxiv.org/abs/hep-ph/9712330}{{\tt
  arXiv:hep-ph/9712330}}.

\bibitem{Fugel:2004ug}
F.~Fugel, B.~A. Kniehl, and M.~Steinhauser, {\em {Two loop electroweak
  correction of O(G(F)M(t)**2) to the Higgs-boson decay into photons}}.
  \href{http://dx.doi.org/10.1016/j.nuclphysb.2004.09.018}{Nucl. Phys. B {\bf
  702} (2004)  333--345}, \href{http://arxiv.org/abs/hep-ph/0405232}{{\tt
  arXiv:hep-ph/0405232}}.

\bibitem{Degrassi:2005mc}
G.~Degrassi and F.~Maltoni, {\em {Two-loop electroweak corrections to the
  Higgs-boson decay H ---\ensuremath{>} gamma gamma}}.
  \href{http://dx.doi.org/10.1016/j.nuclphysb.2005.06.027}{Nucl. Phys. B {\bf
  724} (2005)  183--196}, \href{http://arxiv.org/abs/hep-ph/0504137}{{\tt
  arXiv:hep-ph/0504137}}.

\bibitem{Aglietti:2004nj}
U.~Aglietti, R.~Bonciani, G.~Degrassi, and A.~Vicini, {\em {Two loop light
  fermion contribution to Higgs production and decays}}.
  \href{http://dx.doi.org/10.1016/j.physletb.2004.06.063}{Phys. Lett. B {\bf
  595} (2004)  432--441}, \href{http://arxiv.org/abs/hep-ph/0404071}{{\tt
  arXiv:hep-ph/0404071}}.

\bibitem{Aglietti:2004ki}
U.~Aglietti, R.~Bonciani, G.~Degrassi, and A.~Vicini, {\em {Master integrals
  for the two-loop light fermion contributions to gg ---\ensuremath{>} H and H
  ---\ensuremath{>} gamma gamma}}.
  \href{http://dx.doi.org/10.1016/j.physletb.2004.09.001}{Phys. Lett. B {\bf
  600} (2004)  57--64}, \href{http://arxiv.org/abs/hep-ph/0407162}{{\tt
  arXiv:hep-ph/0407162}}.

\bibitem{Passarino:2007fp}
G.~Passarino, C.~Sturm, and S.~Uccirati, {\em {Complete Two-Loop Corrections to
  H ---\ensuremath{>} gamma gamma}}.
  \href{http://dx.doi.org/10.1016/j.physletb.2007.09.002}{Phys. Lett. B {\bf
  655} (2007)  298--306}, \href{http://arxiv.org/abs/0707.1401}{{\tt
  arXiv:0707.1401 [hep-ph]}}.

\bibitem{Actis:2008ts}
S.~Actis, G.~Passarino, C.~Sturm, and S.~Uccirati, {\em {NNLO Computational
  Techniques: The Cases H ---\ensuremath{>} gamma gamma and H ---\ensuremath{>}
  g g}}. \href{http://dx.doi.org/10.1016/j.nuclphysb.2008.11.024}{Nucl. Phys. B
  {\bf 811} (2009)  182--273}, \href{http://arxiv.org/abs/0809.3667}{{\tt
  arXiv:0809.3667 [hep-ph]}}.

\bibitem{Hahn:2000kx}
T.~Hahn, {\em {Generating Feynman diagrams and amplitudes with FeynArts 3}}.
  \href{http://dx.doi.org/10.1016/S0010-4655(01)00290-9}{Comput. Phys. Commun.
  {\bf 140} (2001)  418--431}, \href{http://arxiv.org/abs/hep-ph/0012260}{{\tt
  arXiv:hep-ph/0012260}}.

\bibitem{Davydychev:1992mt}
A.~I. Davydychev and J.~B. Tausk, {\em {Two loop selfenergy diagrams with
  different masses and the momentum expansion}}.
  \href{http://dx.doi.org/10.1016/0550-3213(93)90338-P}{Nucl. Phys. B {\bf 397}
  (1993)  123--142}.

\bibitem{Degrassi:2010eu}
G.~Degrassi and P.~Slavich, {\em {NLO QCD bottom corrections to Higgs boson
  production in the MSSM}}.
  \href{http://dx.doi.org/10.1007/JHEP11(2010)044}{JHEP {\bf 11} (2010)  044},
  \href{http://arxiv.org/abs/1007.3465}{{\tt arXiv:1007.3465 [hep-ph]}}.

\bibitem{Staub:2008uz}
F.~Staub, {\em {SARAH}}. \href{http://arxiv.org/abs/0806.0538}{{\tt
  arXiv:0806.0538 [hep-ph]}}.

\bibitem{Staub:2009bi}
F.~Staub, {\em {From Superpotential to Model Files for FeynArts and
  CalcHep/CompHep}}. \href{http://dx.doi.org/10.1016/j.cpc.2010.01.011}{Comput.
  Phys. Commun. {\bf 181} (2010)  1077--1086},
  \href{http://arxiv.org/abs/0909.2863}{{\tt arXiv:0909.2863 [hep-ph]}}.

\bibitem{Staub:2010jh}
F.~Staub, {\em {Automatic Calculation of supersymmetric Renormalization Group
  Equations and Self Energies}}.
  \href{http://dx.doi.org/10.1016/j.cpc.2010.11.030}{Comput. Phys. Commun. {\bf
  182} (2011)  808--833}, \href{http://arxiv.org/abs/1002.0840}{{\tt
  arXiv:1002.0840 [hep-ph]}}.

\bibitem{Staub:2012pb}
F.~Staub, {\em {SARAH 3.2: Dirac Gauginos, UFO output, and more}}.
  \href{http://dx.doi.org/10.1016/j.cpc.2013.02.019}{Comput. Phys. Commun. {\bf
  184} (2013)  1792--1809}, \href{http://arxiv.org/abs/1207.0906}{{\tt
  arXiv:1207.0906 [hep-ph]}}.

\bibitem{Staub:2013tta}
F.~Staub, {\em {SARAH 4 : A tool for (not only SUSY) model builders}}.
  \href{http://dx.doi.org/10.1016/j.cpc.2014.02.018}{Comput. Phys. Commun. {\bf
  185} (2014)  1773--1790}, \href{http://arxiv.org/abs/1309.7223}{{\tt
  arXiv:1309.7223 [hep-ph]}}.

\bibitem{BhupalDev:2014bir}
P.~S. Bhupal~Dev and A.~Pilaftsis, {\em {Maximally Symmetric Two Higgs Doublet
  Model with Natural Standard Model Alignment}}.
  \href{http://dx.doi.org/10.1007/JHEP12(2014)024}{JHEP {\bf 12} (2014)  024},
  \href{http://arxiv.org/abs/1408.3405}{{\tt arXiv:1408.3405 [hep-ph]}}.
  [Erratum: JHEP 11, 147 (2015)].

\bibitem{Barroso:2013awa}
A.~Barroso, P.~M. Ferreira, I.~P. Ivanov, and R.~Santos, {\em {Metastability
  bounds on the two Higgs doublet model}}.
  \href{http://dx.doi.org/10.1007/JHEP06(2013)045}{JHEP {\bf 06} (2013)  045},
  \href{http://arxiv.org/abs/1303.5098}{{\tt arXiv:1303.5098 [hep-ph]}}.

\bibitem{Grinstein:2015rtl}
B.~Grinstein, C.~W. Murphy, and P.~Uttayarat, {\em {One-loop corrections to the
  perturbative unitarity bounds in the CP-conserving two-Higgs doublet model
  with a softly broken $ {\mathrm{\mathbb{Z}}}_2 $ symmetry}}.
  \href{http://dx.doi.org/10.1007/JHEP06(2016)070}{JHEP {\bf 06} (2016)  070},
  \href{http://arxiv.org/abs/1512.04567}{{\tt arXiv:1512.04567 [hep-ph]}}.

\bibitem{Cacchio:2016qyh}
V.~Cacchio, D.~Chowdhury, O.~Eberhardt, and C.~W. Murphy, {\em {Next-to-leading
  order unitarity fits in Two-Higgs-Doublet models with soft $\mathbb{Z}_2$
  breaking}}. \href{http://dx.doi.org/10.1007/JHEP11(2016)026}{JHEP {\bf 11}
  (2016)  026}, \href{http://arxiv.org/abs/1609.01290}{{\tt arXiv:1609.01290
  [hep-ph]}}.

\bibitem{Bechtle:2013xfa}
P.~Bechtle, S.~Heinemeyer, O.~St\r{a}l, T.~Stefaniak, and G.~Weiglein, {\em
  {$HiggsSignals$: Confronting arbitrary Higgs sectors with measurements at the
  Tevatron and the LHC}}.
  \href{http://dx.doi.org/10.1140/epjc/s10052-013-2711-4}{Eur. Phys. J. C {\bf
  74} (2014) no.~2, 2711}, \href{http://arxiv.org/abs/1305.1933}{{\tt
  arXiv:1305.1933 [hep-ph]}}.

\bibitem{Bechtle:2020uwn}
P.~Bechtle, S.~Heinemeyer, T.~Klingl, T.~Stefaniak, G.~Weiglein, and
  J.~Wittbrodt, {\em {HiggsSignals-2: Probing new physics with precision Higgs
  measurements in the LHC 13 TeV era}}.
  \href{http://dx.doi.org/10.1140/epjc/s10052-021-08942-y}{Eur. Phys. J. C {\bf
  81} (2021) no.~2, 145}, \href{http://arxiv.org/abs/2012.09197}{{\tt
  arXiv:2012.09197 [hep-ph]}}.

\bibitem{Bechtle:2008jh}
P.~Bechtle, O.~Brein, S.~Heinemeyer, G.~Weiglein, and K.~E. Williams, {\em
  {HiggsBounds: Confronting Arbitrary Higgs Sectors with Exclusion Bounds from
  LEP and the Tevatron}}.
  \href{http://dx.doi.org/10.1016/j.cpc.2009.09.003}{Comput. Phys. Commun. {\bf
  181} (2010)  138--167}, \href{http://arxiv.org/abs/0811.4169}{{\tt
  arXiv:0811.4169 [hep-ph]}}.

\bibitem{Bechtle:2011sb}
P.~Bechtle, O.~Brein, S.~Heinemeyer, G.~Weiglein, and K.~E. Williams, {\em
  {HiggsBounds 2.0.0: Confronting Neutral and Charged Higgs Sector Predictions
  with Exclusion Bounds from LEP and the Tevatron}}.
  \href{http://dx.doi.org/10.1016/j.cpc.2011.07.015}{Comput. Phys. Commun. {\bf
  182} (2011)  2605--2631}, \href{http://arxiv.org/abs/1102.1898}{{\tt
  arXiv:1102.1898 [hep-ph]}}.

\bibitem{Bechtle:2013wla}
P.~Bechtle, O.~Brein, S.~Heinemeyer, O.~St\r{a}l, T.~Stefaniak, G.~Weiglein,
  and K.~E. Williams, {\em {$\mathsf{HiggsBounds}-4$: Improved Tests of
  Extended Higgs Sectors against Exclusion Bounds from LEP, the Tevatron and
  the LHC}}. \href{http://dx.doi.org/10.1140/epjc/s10052-013-2693-2}{Eur. Phys.
  J. C {\bf 74} (2014) no.~3, 2693}, \href{http://arxiv.org/abs/1311.0055}{{\tt
  arXiv:1311.0055 [hep-ph]}}.

\bibitem{Bechtle:2020pkv}
P.~Bechtle, D.~Dercks, S.~Heinemeyer, T.~Klingl, T.~Stefaniak, G.~Weiglein, and
  J.~Wittbrodt, {\em {HiggsBounds-5: Testing Higgs Sectors in the LHC 13 TeV
  Era}}. \href{http://dx.doi.org/10.1140/epjc/s10052-020-08557-9}{Eur. Phys. J.
  C {\bf 80} (2020) no.~12, 1211}, \href{http://arxiv.org/abs/2006.06007}{{\tt
  arXiv:2006.06007 [hep-ph]}}.

\bibitem{Bahl:2021yhk}
H.~Bahl, V.~M. Lozano, T.~Stefaniak, and J.~Wittbrodt, {\em {Testing exotic
  scalars with HiggsBounds}}.
  \href{http://dx.doi.org/10.1140/epjc/s10052-022-10446-2}{Eur. Phys. J. C {\bf
  82} (2022) no.~7, 584}, \href{http://arxiv.org/abs/2109.10366}{{\tt
  arXiv:2109.10366 [hep-ph]}}.

\bibitem{Haller:2018nnx}
J.~Haller, A.~Hoecker, R.~Kogler, K.~M\"onig, T.~Peiffer, and J.~Stelzer, {\em
  {Update of the global electroweak fit and constraints on two-Higgs-doublet
  models}}. \href{http://dx.doi.org/10.1140/epjc/s10052-018-6131-3}{Eur. Phys.
  J. C {\bf 78} (2018) no.~8, 675}, \href{http://arxiv.org/abs/1803.01853}{{\tt
  arXiv:1803.01853 [hep-ph]}}.

\bibitem{Aiko:2023nqj}
M.~Aiko, J.~Braathen, and S.~Kanemura, {\em {Leading two-loop corrections to
  the Higgs di-photon decay in the Inert Doublet Model}}.
  \href{http://arxiv.org/abs/2307.14976}{{\tt arXiv:2307.14976 [hep-ph]}}.

\bibitem{Braathen:2016cqe}
J.~Braathen and M.~D. Goodsell, {\em {Avoiding the Goldstone Boson Catastrophe
  in general renormalisable field theories at two loops}}.
  \href{http://dx.doi.org/10.1007/JHEP12(2016)056}{JHEP {\bf 12} (2016)  056},
  \href{http://arxiv.org/abs/1609.06977}{{\tt arXiv:1609.06977 [hep-ph]}}.

\bibitem{Braathen:2018htl}
J.~Braathen, M.~D. Goodsell, and P.~Slavich, {\em {Matching renormalisable
  couplings: simple schemes and a plot}}.
  \href{http://dx.doi.org/10.1140/epjc/s10052-019-7093-9}{Eur. Phys. J. C {\bf
  79} (2019) no.~8, 669}, \href{http://arxiv.org/abs/1810.09388}{{\tt
  arXiv:1810.09388 [hep-ph]}}.

\bibitem{Hahn:1998yk}
T.~Hahn and M.~Perez-Victoria, {\em {Automatized one loop calculations in
  four-dimensions and D-dimensions}}.
  \href{http://dx.doi.org/10.1016/S0010-4655(98)00173-8}{Comput. Phys. Commun.
  {\bf 118} (1999)  153--165}, \href{http://arxiv.org/abs/hep-ph/9807565}{{\tt
  arXiv:hep-ph/9807565}}.

\end{thebibliography}\endgroup

\end{document}